\documentclass{article}
\usepackage{amsmath}
\usepackage{amsfonts}
\usepackage[hidelinks]{hyperref}
\usepackage{datetime}
\usepackage{cite}

\numberwithin{equation}{section}
\makeatletter
\renewcommand{\@seccntformat}[1]{%
  \csname the#1\endcsname.\ }
\makeatother

\def\p{\partial}
\def\O{\mathcal{O}}
\def\L{\mathcal{L}}
\def\a{\alpha}
\def\b{\beta}

\def\r{\rightarrow}

\def\H{\mathcal{H}}
\def\P{\mathcal{P}}
\def\Q{\mathcal{Q}}
\def\J{\mathcal{J}}

\def\l{\lambda}
\def\d{\delta}
\def\s{\sigma}
\def\ts{\tilde{\sigma}}
\def\e{\epsilon}
\def\om{\omega}

\def\D{\Delta}
\def\Dl{\mathcal{D}_\lambda}

\newcommand{\be}{\begin{equation}}
\newcommand{\ee}{\end{equation}}
\newcommand{\bea}{\begin{eqnarray}}
\newcommand{\eea}{\end{eqnarray}}
\newcommand{\bi}{\begin{itemize}}
\newcommand{\ei}{\end{itemize}}

\title{Symmetries versus the spectrum of $ J\bar T$ - deformed CFTs \vspace{5mm} }

\author{
Monica Guica \vspace{2mm} \\
\\\vspace{2mm}
\emph{\normalsize Institut de Physique Th\'eorique, CEA Saclay, CNRS, 91191 Gif-sur-Yvette, France}\\
\emph{\normalsize Department of Physics, Stockholm University,
AlbaNova, 106 91 Stockholm, Sweden} \vspace{2mm} \\ 
\emph{\normalsize Nordita, Roslagstullsbacken 23, SE-106 91 Stockholm, Sweden} \vspace{2mm}}

\date{}
\begin{document}

\maketitle

\begin{abstract}
\vskip 2mm

\noindent It has been recently shown that classical $J\bar T$ - deformed CFTs possess an infinite-dimensional Witt-Ka\v{c}-Moody symmetry, generated by certain field-dependent coordinate  and gauge transformations. On a cylinder, however, the equal spacing of the descendants' energies predicted by such a symmetry algebra is inconsistent with the known finite-size spectrum of $J\bar T$ - deformed CFTs. Also, the associated quantum symmetry generators do not have a proper action on the Hilbert space. In this article, we resolve this tension by finding a new set of (classical) conserved charges, whose action is consistent with semiclassical quantization, and which are related to the previous symmetry generators by a type of energy-dependent spectral flow. The previous inconsistency between the algebra and the spectrum is resolved because the energy  operator does not belong to the spectrally flowed sector. 
\end{abstract}

\tableofcontents

\newpage

\section{Introduction and statement of the problem}

The discovery of the AdS/CFT correspondence \cite{Maldacena:1997re} has marked a major step in our current understanding of quantum gravity.  While there are good reasons to believe that gravity in general backgrounds is holographic, various clues point towards the fact that for asymptotically flat spacetimes \cite{Li:2010dr} or spacetimes related to the near-horizon geometry of extremal black holes\cite{ElShowk:2011cm}, the dual QFT may be non-local. However, non-local quantum field theories are still relatively poorly understood, in comparison with their local counterparts. 

In \cite{Smirnov:2016lqw}, Smirnov and Zamolodchikov (see also \cite{Cavaglia:2016oda}) laid out the construction of a set of tractable irrelevant deformations of two-dimensional local QFTs which result into QFTs that are non-local, yet appear to be UV complete \cite{Dubovsky:2012wk,Dubovsky:2013ira}.  Moreover, these theories are solvable, in the sense that one can compute their spectrum, $S$-matrix and other observables \cite{Cardy:2018sdv,Dubovsky:2018bmo,Cardy:2019qao,Kruthoff:2020hsi} in terms of the corresponding quantities in the undeformed QFT. Even more interestingly, certain single-trace analogues of the Smirnov-Zamolodchikov deformations have been related to holography in non-asymptotically AdS spacetimes \cite{Giveon:2017nie,Apolo:2018qpq,Chakraborty:2018vja}. 

The  Smirnov-Zamolodchikov deformations are constructed from bilinears of two conserved currents. The best studied of these is the so-called $T\bar T$ deformation, constructed from the components of the stress tensor. Deformations constructed from a $U(1)$ current and the stress tensor, such as the $J\bar T$ \cite{Guica:2017lia} and the $JT_a$ \cite{Anous:2019osb} deformations are also relatively well studied. Of these, the $J\bar T$ deformation of two-dimensional CFTs is the simplest, as the non-locality of the deformed QFT is concentrated exclusively to the right-moving side, and the theory stays local and conformal on the left. The effect of performing several of these deformations simultaneously has been studied in e.g. \cite{Frolov:2019xzi,LeFloch:2019rut}. 

It is interesting to better understand  the structure of the Smirnov-Zamolodchikov deformations from a quantum-field-theoretical point of view. It has been recently shown \cite{Guica:2020uhm} that at least at the classical level, $T\bar T$, $J\bar T$ and $JT_a$ - deformed CFTs all posses an infinite-dimensional set of field-dependent symmetries, whose algebra consists of two commuting copies of the  Witt or the  Witt-Ka\v{c}-Moody algebra, if $U(1)$ currents are present. This structure was suggested by the previous holographic analyses of \cite{Guica:2019nzm} for $T\bar T$ and \cite{Bzowski:2018pcy} for $J\bar T$. These analyses also allowed for the computation of the central extension of the symmetry algebra, which becomes Virasoro-Ka\v{c}-Moody.  If these symmetries survive quantization, then we would conclude that  $T\bar T$, $J\bar T$ and $JT_a$ - deformed CFTs correspond to a non-local version of two-dimensional CFTs, with a similarly rigid structure that would highly deserve  further exploration. 

There is, however, a problem, which can be seen already at the semiclassical level. The symmetry analysis of \cite{Guica:2020uhm} is valid on both the plane and the cylinder. In the latter case, one immediately encounters a tension between the equally-spaced energies 
of the  Virasoro descendants predicted by the symmetry analysis and the energies  of the deformed eigenstates in $T\bar T$, $J\bar T$ and $JT_a$ - deformed CFTs, which  take a square root form. In this note, we address this issue for the simplest case of the $J\bar T$ deformation, where the locality of the left-moving side provides a  useful guiding principle for finding its resolution. 

To state the problem explicitly, we start with a review of the relevant facts. The finite-size energy spectrum of $J\bar T$ - deformed CFTs placed on a cylinder of circumference $R$ is given by \cite{Chakraborty:2018vja,Bzowski:2018pcy} % \emph{Correct! Notation?}

\be
E_R \equiv \frac{E-P}{2} = \frac{2}{\l^2} \left( R-\l J_0 -\sqrt{(R-\l J_0)^2 - \l^2 E_R^{(0)}}\right) \;, \;\;\;\;\; E_L= E_R+P \label{defeng}
\ee
where $\l$ is the deformation parameter (with dimensions of length), $J_0$ is the left-moving charge in the undeformed CFT, $P$ is the  quantized momentum,  $E^{(0)}_R$
is the undeformed right-moving energy and  the chiral anomaly coefficient  has been set to $2\pi$. Note that while in the undeformed CFT, the energies of  the right-moving (Virasoro and Ka\v{c}-Moody) descendats of a primary state are  equally-spaced, those of the corresponding $J\bar T$-deformed   descendants are not. The energies of the left-moving descendants do have equal spacing, since they are obtained by raising $E_L = E_R+P$ with $E_R^{(0)}$ and $J_0$ held fixed. 

We note in passing that the relation between the undeformed and deformed energies can be suggestively written as spectral flow \cite{Schwimmer:1986mf}, with a parameter $\l E_R$  proportional to the right-moving energy  %\emph{Check conventions!}

\be
 E_L R = E_L^{(0)} R + \l J_0 E_R + \frac{\l^2}{4} E_R^2 \;, \;\;\;\;\;\;\; E_R(R-\l w) = E_R^{(0)} R + \l \bar J_0 E_R + \frac{\l^2}{4} E_R^2 \label{specasspfl}
\ee
Here, $\bar J_0$ is the right-moving $U(1)$ charge in the undeformed CFT, and $w = J_0 - \bar J_0$ is the winding charge. This  observation  will be quite useful later. It has already been used in deriving the spectrum  in  presence of a chiral anomaly \cite{Chakraborty:2018vja} and for organising the conformal dimensions of $J\bar T$-deformed CFTs on the plane \cite{Guica:2019vnb}. 

The symmetries of $J\bar T$ - deformed CFTs consist of, first,  an infinite set of left-moving conformal and $U(1)$ gauge symmetries that enhance the $SL(2,\mathbb{R})_L \times U(1)_L$ global symmetries of the theory. These symmetries are parametrized by two  arbitrary functions of the left-moving coordinate, $U=\s + t$.  In the general Hamiltonian framework for $J\bar T$ - deformed CFTs developed in \cite{Guica:2020uhm}, they are generated by

\be
Q_f = \int d\s \, f(U/R)\, \H_L  \;, \;\;\;\;\;\;\; P_\eta = \int d\s\,  \eta (U/R) \, (\J_+ + \frac{\l}{2} \H_R) \label{lmch}
\ee
where, in order for the argument of the functions to have periodicity one, the coordinate $U$ has been divided by the  circumference of the circle.  $\H_L= \H_R + \P$ is the left-moving Hamiltonian density, where the right-moving Hamiltonian density $\H_R$ is given in terms of its undeformed counterpart $\H_R^{(0)}$ by a formula entirely analogous to \eqref{defeng}, with $J_0$ replaced by  $\J_+$,  the left-moving current density. The commutation relations of the deformed generators are then fixed by those of the undeformed currents, and one can show that the left-moving charge algebra is precisely Witt-Ka\v{c}-Moody %\emph{Check!}

\be
\{ Q_f, Q_g \} = \frac{1}{R}\, Q_{fg'-f' g} \;, \;\;\;\;\;\;\{ Q_f, P_\eta\} = \frac{1}{R}\, P_{f \eta'} \;, \;\;\;\;\;\;\{ P_\eta, P_\chi\} = \frac{1}{2} \int d\s \chi \p_\s \eta
\ee
%The factor of $1/R$ is due to the fact that we need to divide the coordinate $U$ by it in order to make the functions periodic. It
The factor of $R$ can be absorbed into a rescaling of the pseudoconformal charges $\bar Q_{\bar f}$. 

The second set of infinite-dimensional symmetries of $J\bar T$ - deformed CFTs are field-dependent, and are generated by functions of the field-dependent coordinate

\be
v = \s - t - \l \phi
\ee
where $\phi$ is related to the current $J$ via $J=\star d \phi$. %\emph{Careful here $J $ vs. $\tilde J$!} 
The conserved pseudo-conformal and U(1) charges are given by 
\be
\bar Q_{\bar f} = \int d\s \, \bar f\left(\frac{v}{R_v}\right)\, \H_R  \;, \;\;\;\;\;\;\; \bar P^{KM}_{\bar \eta} = \int d\s\,  \bar \eta \left(\frac{v}{R_v}\right) \, (\J_- + \frac{\l}{2} \H_R) \label{rmch}
\ee
where $R_v = R - \l w$ is the field-dependent radius of the field-depedent coordinate $v$, %, where $w = J_0 - \bar J_0$ is the winding charge of the field configuration. \emph{Careful, check!} 
and the particular combination $\bar P^{KM}$ of the right-moving currents is singled out by its simple commutation relations. 
Remarkably, these charges entirely commute with the left-moving ones, and the charge algebra is still a functional Witt-Ka\v{c}-Moody algebra 
\be
\{ \bar Q_{\bar f}, \bar Q_{\bar g}\} =\frac{1}{R_v} \bar Q_{\bar f' \bar g - \bar f \bar g'} \;, \;\;\;\;\; \{ \bar Q_{\bar f}, \bar P_{\bar \eta}^{KM}\} = -\frac{1}{R_v} \bar P^{KM}_{\bar f \bar \eta'} \;, \;\;\;\;\;\; \{ \bar P^{KM}_{\bar \chi}, \bar P_{\bar \eta}^{KM}\} = -\frac{1}{2} \int d\s \bar \chi \p_\s \bar \eta  \label{rmchalg}
\ee
The word `functional' above refers to the fact that the factors of $R_v$ can be absorbed into a redefinition of the $'$ to mean `derivative with respect to $v$', rather than to the full rescaled argument of the functions $\bar f, \bar \eta$, as used above. In any case, when labeling the charges in terms of the Fourier modes of $\bar f, \bar \eta$, etc., the field-dependent radius will explicitly appear  in the algebra\footnote{In the $J\bar T$ case, we can simply rescale the generators by $R_v$ to obtain a usual Witt-Ka\v{c}-Moody  algebra.%, unlike in the $T\bar T$ case, but we will not do so.}
}.  While this algebra is not exactly Witt-Ka\v{c}-Moody, it still predicts, upon quantization,  an equally-spaced spectrum of descendants, which is incompatible with the energy formula \eqref{defeng}. 

It is in fact not hard to notice  already from their classical Poisson brackets, that the right-moving charges \eqref{rmch} will not have a proper action on the semiclassical phase space of the theory, where the charges associated to the global $U(1)$ symmetry and the momentum are quantized.  Concretely, the problem appears to lie in the commutators of the right-moving generators with the $U(1)$ charges

\be
\{ \bar Q_{\bar f}, J_0 \} = \frac{1}{R_v} \int_0^R d \s  \bar f' \left(\frac{v}{R_v} \right) \H_R (\s) \{ v(\s), J_0 \}   = - \frac{\l}{2 R_v}\, \bar Q_{\bar f'}=\{ \bar Q_{\bar f}, \bar J_0 \}
\ee
and the analogous commutators of the right-moving $U(1)$ generators $\bar P_{\bar \eta}$ with $J_0 $ and $\bar J_0$. What this means is that  $J_0 + \bar J_0$, which represents the global $U(1)$   charge of the configuration and which should be quantized, is changed by a non-integer amount (more precisely, $2\pi \l n/R_v$, with $n \in \mathbb{Z}$) by the action of the semiclassically quantized right-moving generators on a state in the deformed theory. A similar statement holds for the momentum, which from  
\eqref{rmchalg} can be shown to satisfy 
%
%
%
%\be
%\{ \bar Q_{\bar f}, \J_\pm^0 \} = \int_0^R d \s d\ts \left[ \bar f (v)\{ \H_R(\s), \J_\pm (\ts)\} + \frac{\bar f'(v)}{R_v} \H_R (\s) \{ \s-\l \phi(\s), \J_\pm (\ts) \}  \right] = - \frac{\l}{2 R_v}\, \bar Q_{\bar f'}
%\ee
\be
%\{ \bar P_{\bar \eta}, \J_\pm^0 \} = - \frac{\l}{2 R_v} \bar P_{\bar \eta'}\;, \;\;\;\;\;\;
\{ \bar Q_{\bar f}, P\} = -\frac{1}{R_v} \bar Q_{\bar f'}\;, \;\;\;\;\;\;\; \{ \bar P_{\bar \eta}, P\} = -\frac{1}{R_v} \bar P_{\bar \eta'}
\ee
i.e. it is changed by units of $2\pi/R_v$, instead of $2\pi/R$. These  observations imply that the action of the right-moving generators (with the exception of the global right-moving energy and charge) on a field configuration  is in tension with  semiclassical quantization. %Also the momentum would be changed by an amount inconsistent with its quantization.
 Hence, the naive quantum versions of the charges \eqref{rmch} do not act properly on the Hilbert space of $J\bar T$ - deformed CFTs on a cylinder. 
 
While having an infinite set of symmetry generators that do not properly act on the Hilbert space of the system is not very useful, an interesting question is whether these generators can be modified in such a way that their algebra is preserved, but their action on the Hilbert space is rectified. In this note, we  show that this is indeed possible, by explicitly constructing   an infinite set of charges that, upon quantization, would  act  on the deformed finite-size Hilbert space in a way consistent with charge and momentum quantization. These charges can therefore  be used to organise the spectrum of the deformed CFT.

To find them, we study the flow equation with respect to the deformation parameter $\l$ of the various  energy eigenstates and compare it to the flow of the symmetry generators \eqref{lmch} and \eqref{rmch}. Introducing a new set of operators that relate deformed descendant states to the deformed primaries, we find that they are related to the previously discussed symmetry generators by a type of energy-dependent spectral flow transformation. The new symmetry generators are conserved and satisfy a Witt-Ka\v{c}-Moody algebra with a field-independent radius.
Their commutation relations with the energy and momentum are non-trivial though, as the latter two operators belong to the unflowed sector. This resolves the apparent tension between the symmetry algebra and the spectrum of $J\bar T$ - deformed CFTs.  
We should note that while our analysis is mostly classical - i.e., at the level of Poisson brackets - the quantum generalization of these generators now appears to be straightforward.

%This implies   that on a circle there is a different relevant symmetry algebra, which has funny commutators with $L_0$ (but Virasoro among themselves - check!).

This paper is organised as follows. In section \ref{flst}, we derive the flow equation satisfied by the energy eigenstates in a $J\bar T$ - deformed CFT, by adapting the method used in \cite{Kruthoff:2020hsi} to study the flow of states under the $T\bar T$ deformation. We subsequently compare this to the flow equations satisfied by the symmetry generators, and argue that the two sets  of generators must be related by a similarity transformation that we denote as ``spectral flow''. In section \ref{flop}, we proceed to finding the flowed operators, first perturbatively and then by making an all-orders proposal, whose consistency we then check. % We conclude with a discussion and future directions. 
The technical details of the very many  Poisson brackets we need are collected in the appendices.

\section{Flow of the eigenstates versus the symmetry generators \label{flst}}

\subsection{The flow of energy eigenstates}

Let $| n_\l \rangle $ be an energy (and momentum, and charge) eigenstate in the theory deformed by an amount $\l$. As  $\l$ is infinitesimally changed, the change in the eigenstate is given by first-order quantum-mechanical perturbation theory

\be
\p_\l |n_\l \rangle = \sum_{m\neq n} \frac{\langle m_\l | \p_\l H| n_\l \rangle}{E_n^\l- E_m^\l} \, |m_\l \rangle \label{fopertth}
\ee
where $\p_\l H$ is the change in the Hamiltonian. For convenience, we take the deforming operator to be $\tilde J \bar T$, rather than $J\bar T$, where $\tilde J = \star d \phi$ is a topologically conserved current. Its components are

\be
\tilde J_t = \phi' \;, \;\;\;\;\;\;\; \tilde J_\s = \p_{\pi} \H
\ee
where $\pi$ is the canonical momentum conjugate to $\phi$. 
One can easily check, using the method developed in \cite{Guica:2020uhm}, that the $\tilde J \bar T$ deformation leads to the same deformed Hamiltonian density as $J \bar T $. %, and we use it because the formulae for the deformed states are also somewhat simplified. \emph{Remember why!} 
Consequently, the change in the Hamiltonian is given by\footnote{We are using the conventions of \cite{Guica:2020uhm} throughout this article.} 

\be
\p_\l H(\l) = -\int d\s \O_{\tilde J\bar T} = -\int d\s \, \e^{\a\b}\tilde  J_\a (\s) T_{\b V} (\s)
\ee
To find the general solution for the deformed eigenstates, we will use the technique proposed by \cite{Kruthoff:2020hsi}. On an equal-time slice, we write

\be
\int d\s \O_{\tilde J\bar T} = \int d\s d\ts \,  \e^{\a\b}\tilde  J_\a (\s) \d(\s-\ts) T_{\b V} (\ts)
\ee
It is useful to introduce the Green's function on the cylinder of circumference $R$ %(taken to be general in this section)

\be
G(\s ) = \frac{1}{2\pi i}\sum_{m\neq 0} \frac{1}{m} e^{2\pi i m \s/R} = \frac{1}{2} \mbox{sgn} (\s) - \frac{\s}{R} \label{greensf}
\ee
which is single-valued and satisfies
\be
\p_\s G(\s-\ts) = \d(\s-\ts) - \frac{1}{R}
\ee
Then, we can rewrite the deforming operator as  
\bea
\int d\s \O_{\tilde J\bar T} &= &\int d\s d\ts \left[ \tilde J_\s (\s) \left(\frac{1}{R} + \p_\s G (\s-\ts)\right) T_{tV} (\ts)- \tilde J_t (\s) \left(\frac{1}{R} - \p_{\ts} G (\s-\ts)\right) T_{\s V} (\ts)\right]  \nonumber \\
&=&  \frac{1}{R} \e^{\a\b} \tilde J^0_\a \, T^0_{\b V} - \int d\s d\ts G(\s-\ts) (\p_\s \tilde J_\s (\s) T_{t V} (\ts) + \tilde J_t (\s) \p_{\ts} T_{\s V} (\ts))  \nonumber \\
&=&  \frac{1}{R} \e^{\a\b} \tilde J^0_\a \, T^0_{\b V} - \p_t \int d\s d\ts G(\s-\ts) \tilde J_t (\s) T_{tV}(\ts) 
\eea
where we introduced the notation 

\be
\tilde J^0_\a \equiv \int d\s \, \tilde J_\a \;, \;\;\;\;\;\;\;T^0_{\b V} \equiv \int d\s \, T_{\b V}
\ee
Naturally, the integral of the time components of the currents above will yield the associated conserved charges, i.e. the winding, $w$, and  respectively (minus) the right-moving energy, $-E_R$. We further use the manipulations of \cite{Kruthoff:2020hsi} to rewrite the regulated denominator of \eqref{fopertth} as an integral, in terms of which the flow equation for the states becomes 
\bea
\p_\l |n_\l \rangle &=& - i \sum_{m\neq n} \int_{-\infty}^0 dt \,  e^{t \e} \, |m_\l \rangle \langle m_\l | \p_\l H_\l (t) |n_\l \rangle  \nonumber \\
&=&  i \sum_{m\neq n}    \,  |m_\l \rangle \langle m_\l \left|\,\frac{1}{R} \int_{-\infty}^0 dt \,  e^{t \e} \e^{\a\b}  \tilde J^0_\a \, T^0_{\b V} - \int d\s d\ts G(\s-\ts) \tilde J_t (\s) T_{tV}(\ts) \right|n_\l \rangle \label{plnint}
\eea
Here, $\e >0$ is an infinitesimal regulator used to make the integral converge, and  the second term is evaluated on the $t=0$ slice. Since we are working on the cylinder, the first term cannot be ignored.  To evaluate it, we need the explicit form of the spatial components of $\tilde J$ and the right-moving translation generator, which can be worked out using the formulae given in \cite{Guica:2020uhm}

\be
\tilde J_\s = \p_\pi\H = \phi' + 2 \p_{\pi} \H_R%= \phi_1' + \frac{\p_\a F}{\l}+2 \J_- \p_x F 
= \phi' + 2 \frac{\J_- + \l \H_R/2}{\sqrt{}}
\ee
and 
\be
T_{\s V} = 2 T_{VV} - \H_R =% \frac{2}{\l^2} \p_\a F - \H_R  = 
2\frac{\H_R}{\sqrt{}} - \H_R
\ee
where the somewhat unusual notation $\sqrt{}\;$ is a shorthand for $\sqrt{(1-\l \J_+)^2 - \l^2 \H_R^{(0)}}$. 
Therefore, 
\be
\e^{\a\b}\tilde J_\a^0 T_{\b V}^0 = - 2 w \int d\s \frac{\H_R}{\sqrt{}} - 2 E_R \int d\s \frac{\J_-+\l \H_R/2}{\sqrt{}}
\ee
In order to perform the time integral in \eqref{plnint}, we would like to rewrite the above operator as a time derivative, i.e. as a commutator with the Hamiltonian. This can be achieved by introducing the zero modes
\be
\phi_0 \equiv \int_0^R d\s \, \phi(\s) \;,\;\;\;\;\;\;\;\chi_0 \equiv \int_0^R  d\s \,\chi(\s)
\ee
where the auxiliary non-local field $\chi$ is defined via

\be
 \p_\s \chi \equiv \H_R
\ee
Such fields also made their appearance in the analysis of the charge algebra for $T\bar T$ - deformed CFTs in \cite{Guica:2020uhm}, though they were not given an explicit name.

The Poisson brackets of the fields $\phi$ and $\chi$ (and, consequently, of their zero modes $\phi_0$ and $\chi_0$) are fixed by the Poisson brackets of the corresponding currents $\J_\pm, \H_R, \P$, up to some possible integration functions. While the choice of these functions is straightforward for the $\phi$ commutators, as $\phi$ is a local field, it is however somewhat subtle for the case of the $\chi$ commutators, because $\chi$ is non-local. In appendix \ref{pbchi}, we perform a rather detailed analysis of the Jacobi identities that constrain these integration functions, with the result 
\be
\{\chi_0, H \} = -2 \int_0^R d\s \frac{\H_R}{\sqrt{}} + E_R + \frac{E_R R}{R_v}\;, \;\;\;\;\;\;\;\{ \phi_0, H\} = 2 \int_0^R d\s \frac{\J_-+\l/2 \H_R}{\sqrt{}} + w
\ee
where $E_R= \int d\s \H_R$ is the right-moving energy operator.  Consequently, we can write 

\be 
\e^{\a\b}\tilde J_\a^0 T_{\b V}^0 = - \{ H,  w \chi_0 - E_R \phi_0 \} - w E_R \frac{R}{R_v} = \frac{d}{dt} ( w \chi_0 - E_R \phi_0 ) - w  \frac{E_R R}{R_v}
\ee
Plugging this into \eqref{plnint}, the last term drops out, as it is evaluated between two different energy-momentum eigenstates. The integral of the first term over the half line yields, in the limit $\e \r 0$ 

\be
\p_\l | n_\l \rangle = i \sum_{m\neq n} |m_\l\rangle \langle m_\l \left|\, \frac{w \chi_0 - E_R \phi_0}{R} + \int d\s d\ts G(\s-\ts) \phi'(\s) \H_R (\ts) \, \right|n_\l\rangle \label{flopsumst}
\ee
%
%\be
%\frac{1}{R}\int_{-\infty}^0 dt e^{t \e} \e^{\a\b}\tilde J_\a^0 T_{\b V}^0 = \frac{1}{R}\int_{-\infty}^0 dt e^{t \e} \left(w \dot \chi_0 + w \{ P, \chi_0\} - E_R \dot \phi_0 - E_R \{ P, \phi_0\} \right) = \frac{w}{ R} \chi_0 - \frac{E_R}{R} \,\phi_0 
%\ee
%up to terms that entirely commute with $H_R/ E_R$. 
We will be denoting these two contributions as $\D \O$ and respectively $\hat \O$, defined as 

\be
\D \O \equiv  \frac{w \chi_0 - E_R \phi_0}{R} \;, \;\;\;\;\;\;\;\; \hat \O \equiv  \int d\s d\ts G(\s-\ts) \phi'(\s) \H_R (\ts) \label{defndelOhatO}
\ee
and their sum will be denoted as $\O_{tot} = \D \O + \hat \O$. If we make use of the identity 

\be
\int d\ts \phi'(\ts) G(\ts-\s) = \left. \phi(\ts) G(\ts-\s)\right|_{0}^R - \phi(\s) + \phi_0 =- \hat \phi (\s) + \phi_0
\ee
where $\hat \phi (\s)= \phi (\s) - w \s/R$ is the scalar field with its winding mode removed (which is thus single-valued on the circle), 
then an alternate expression for $\O_{tot}$ is 
\be
\O_{tot} = \frac{w \chi_0}{R} - \int d\s \hat \phi(\s) \H_R
\ee
which is rather useful in computing its Poisson brackets. 

As a final step of our manipulations, we use the assumed completeness of the set of states to rewrite the flow equation for the energy eigenstates as 

\be
\p_\l | n_\l \rangle = i \O_{tot}  |n_\l\rangle - i | n_\l \rangle \langle n_\l | \O_{tot}  |n_\l\rangle
\ee
Introducing an operator, $D$, which is diagonal  in the energy eigenbasis and whose matrix elements are defined as  $\langle n_\l | D | n_\l \rangle = \langle n_\l  | \O_{tot} | n_\l  \rangle $, we can  rewrite the flow equation for the eigenstates in its final form 

\be
\boxed{\p_\l | n_\l \rangle = i (\O_{tot}-D) | n_\l \rangle } \label{flowstates}
\ee
Thus, to understand the flow of the states, we need to understand also which parts of $\O_{tot}$ have non-zero expectation values in the energy eigenstates. This is a quite non-trivial task for arbitrary values of the flow parameter. We can nevertheless attempt to understand this problem perturbatively.  For example, at $\l=0$, we can use \eqref{greensf} to evaluate 

\be
i \hat \O =  \sum_{m\neq 0} \frac{1}{m} :\tilde J_m \bar L_m :+ \ldots
\ee 
where $\tilde J_m = J_m - \bar J_{-m}$ and $\bar L_m$ are the Fourier modes of $\phi'$ and respectively $\H_R$ in the undeformed CFT, and the colons denote normal ordering. Since the sum is strictly over non- zero modes, it is clear that 
the  expectation value  of this operator in any energy eigenstate of the undeformed CFT is zero.
%\emph{ Off by an overall minus sign?? Careful definition $\bar L_m$!}
 Thus, $\hat \O$ does not contribute to $D$, at least at $\l=0$. On the other hand,  the expectation value of $\D \O$ vanishes between any two different energy eigenstates at $\l=0$,  as one can see by evaluating %\emph{Careful $\chi_0$!}

\be
\langle m |[H, \Delta \O]| n\rangle = (E_m-E_n) \langle m| \Delta \O | n \rangle =  i \langle m| (J_0 + \bar J_0) E_R | n \rangle =0
\ee
which implies that\footnote{For degenerate eigenstates, one can repeat the argument for the commutator with  other globally conserved charges.  }  $\langle m| \Delta \O | n\rangle =0\;,\;$ $\forall m\neq n$. Thus, we find that at $\l=0$, $D= \D \O$. 

%for $E_m \neq E_n$. Other commutators can be evaluated to distinguish between degenerate eigenstates. 

 At higher orders in perturbation theory, $\D \O$ may start having non-zero matrix elements between different eigenstates, which would therefore not contribute to $D$.  To understand what happens, we should
 study the change with $\l$ of   the matrix elements $\langle m | \Delta \O |n\rangle$ %\emph{Why $\D \O$ and not $\O_{tot}$?}

\be
\p_\l \langle m | \Delta \O |n\rangle= \langle m | \p_\l \D \O - i [\O_{tot}-D, \D \O] |n\rangle =  \langle m |\hat{\mathcal{D}}_\l \D \O | n \rangle+ i (D_m-D_n) \langle m | \Delta \O |n\rangle
\ee
where the flow operator $\hat{\mathcal{D}}_\l$ is defined as %\emph{Inconsistent sign map PB to commutators?}
\be
\hat{\mathcal{D}}_\l \equiv \p_\l -i [\O_{tot},\; \cdot\;]
\ee 
Using the explicit expression, \eqref{DlD0}, for $\mathcal{D}_\l \D \O$ computed in the next section, we see that at $\l =0$, the only contribution to $\langle n | \mathcal{D}_\l \D \O | n \rangle$ comes from the terms proportional to the zero modes of the fields $\phi$ and $\chi$. The $\l$ dependence of the diagonal matrix elements of $\hat \O$ can be studied by plugging in the known expression for $\H_R(\l)$. At first order in $\l$, it also does not look like  this operator has  non-zero diagonal matrix elements in the energy eigenbasis\footnote{Using the formulae in appendix \ref{flowapp}, it can be shown that  $\hat \O$ satisfies the very simple flow equation $\Dl \hat \O = w \,\hat \O/ R_v$ (for $a=-1$), which can be used to  evaluate its contribution, if any, at higher orders in $\l$.}, and thus it will not contribute to $D$. %\emph{Can we study its flow equation to  prove this?}

To summarize, up to first order in $\l$, we expect that

\be
D = \D \O - \l (\mathcal{D}_\l \D \O)_{no \, z.m.}  + \O(\l^2) \label{Dpert}
\ee
i.e., we are subtracting all the off-diagonal contributions to $\D \O$ up to this order. Performing this analysis to higher order looks increasingly cumbersome, and we may need a better method. 

The discussion so far holds for states defined on the $t=0$ slice. It is interesting to also consider the flow equation for states defined at a time  $t$ instead of $t=0$. Our derivation of  the flow operator \eqref{flopsumst} still holds, except that it should now be evaluated at time $t$, rather than $t=0$. 
%
% but we now need to evaluate $\D\O$ and $\hat \O$ defined in \eqref{defndelOhatO} at time $t$. These  operators are obtained as usual by conjugating their $t=0$ counterparts with the evolution operator.
  Since the states at $t$ are related to the states at $t=0$ by a $\l$ - dependent energy factor, the flow equation is best written as%/acquires and extra term \emph{Check and explain!}

\be
\p_\l | n_\l (t) \rangle =  i (\O_{tot} (t) - \p_\l E_n \, t - D (t)) | n_\l (t)\rangle \;, \;\;\;\;\;\; \p_\l E = 2 \frac{E_R Q_K}{R-\l Q_K}  \label{tdepflow}
\ee
where $Q_K \equiv J_0 + \l E_R/2$, and the expression for $\p_\l E $ is obtained from \eqref{defeng}.  The matrix elements of the operator $D(t)$ are defined as the expectation values of $\O_{tot}(t) $ in energy eigenstates.

\subsection{Flow of the symmetry generators}

Having understood the flow of the energy eigenstates with respect to $\l$, we would now like to discuss the corresponding flow of the symmetry generators $Q_f, P_\eta, \bar Q_{\bar f}$ and $\bar P_{\bar \eta}$. It is useful to compute the action of the operator $\Dl$ defined above on these generators. Our analysis will be classical, and thus we will be using the Poisson bracket counterpart of this flow operator, i.e

\be
\mathcal{D}_\l \equiv \p_\l +\{\O_{tot},\; \cdot\;\} \label{dldef}
\ee 
obtained by the usual replacement $[\;,\;] \r i \{\;, \;\}$. 
 To compute the action of $\Dl$, we will need the Poisson brackets of the various currents in the $J\bar T$-deformed CFT, which were derived in \cite{Guica:2020uhm} and are collected for convenience in appendix \ref{pbapp}. We will also need the Poisson brackets of the various symmetry currents with the zero modes of $\chi$ and $\phi$. 
 
  The commutators of the zero mode of $\phi$ are obtained by simply integrating the corresponding commutators of the field $\phi(\s)$, and we obtain

\be
\{  \phi_0, \H_R(\s)\} =  \frac{\J_- + \frac{\l}{2} \H_R}{\sqrt{(1-\l \J_+)^2 - \l^2 \H_R^{(0)}}}\;, \;\;\;\;\;\; \{ \phi_0, \P(\s)\} =  \phi'(\s)\;, \;\;\;\;\;\;\{ \phi_0, \J_\pm\} =  \frac{1}{2}
\ee
Note that in the CFT limit, $\{ \phi_0, \H_L\} = \J_+$ and $\{ \phi_0, \H_R\} = \J_-$, so the exponential of this operator is precisely what generates spectral flow for the left- and the right-movers. 

The Poisson brackets of the zero mode $\chi_0$ are significantly more involved, due to the fact that the ancillary field $\chi$ is non-local, being defined as the integral of the local current $\H_R$. Consequently, its Poisson brackets are defined only up to certain integration functions, whose form is non-trivially constrained by various Jacobi identities. These constraints are analysed in detail in appendix \ref{pbchi}, and the end result for the various commutators of $\chi_0$ is
\bea
\{ \chi_0, \J_+\} &=& - \frac{\l R}{2} \p_\s \left[\frac{\H_R}{\sqrt{}}\left(1+a - \frac{\l\hat\phi}{R_v} \right) \right]- \frac{\l R}{2 R_v} \H_R \\
  \{ \chi_0, \J_-\} &=& R \, \p_\s \left[\frac{\J_-}{\sqrt{}}\left(1+a - \frac{\l\hat\phi}{R_v}\right) \right] - \frac{\l R}{2 R_v} \H_R \\
   \{ \chi_0, \H_R\} & = & - \frac{\H_R}{\sqrt{}}+ \frac{R}{R_v} \H_R + R  \p_\s \left(\frac{\H_R}{\sqrt{}}\left(1+a - \frac{\l\hat\phi}{R_v} \right)\right) \\
    \{ \chi_0, \P\} &= &\H_R - \frac{R}{R_v} \H_R- R  \p_\s \left(\frac{\H_R}{\sqrt{}}\left(1+a  - \frac{\l\hat\phi}{R_v}  \right) \right)  \\
\{\chi_0, \phi\} &=& %- R \frac{\J_-+\l \H_R/2}{\sqrt{}}  \left(1+\frac{v+t}{R_v} - \frac{\s}{R} \right)=
-  \frac{\J_-+\l \H_R/2}{\sqrt{}} R  \left(1+a-\frac{\l \hat \phi}{R_v}\right)
\eea
where, as before,  $\hat \phi = \phi - w \s/R$ equals $\phi$ with its winding mode removed. The terms  proportional to the constant $a$  are allowed by all the Jacobi identities we have studied\footnote{This does not mean that there cannot exist other Jacobi identities that constrain the value of $a$, or that require the introduction of new terms in the commutators above.  Our analysis is thus valid up to this caveat.}. Since its value does not seem to be fixed and, moreover, it drops out from most of our subsequent computations, we will henceforth fix it to the convenient value $a=-1$.

 Using these, one can compute the flow equations for the various currents, which are spelled out for convenience in appendix \ref{flowapp}, and from them  derive the flow of the conserved charges. One finds that the left-moving charges  are simply constant with respect to $\Dl$

\be
\Dl Q_f = \Dl P_\eta =0
\ee
while the right-moving ones satisfy
\be
\Dl \bar Q_{\bar f} = \frac{w}{R_v} \bar Q_{\bar f} - \frac{w t}{R_v^2}  \bar Q_{\bar f'}\;, \;\;\;\;\;\;\; \Dl \bar P^{KM}_{\bar \eta} =  - \frac{w t}{R_v^2} \bar P^{KM}_{\bar \eta'} \label{flowrmch}
\ee
Note that the first term on the right-hand side of the $\bar Q_{\bar f}$ flow is necessary in order  for the flow equation to be compatible with the charge algebra \eqref{rmchalg}, as the latter contains an explicit factor of $1/R_v$, whose $\l$ derivative does not vanish. If we consider instead the rescaled dimensionless charges $R_v \bar Q_{\bar f}$,  they satisfy a flow equation analogous to that of $\bar P^{KM}$. Their algebra is also the standard Witt-Ka\v{c}-Moody algebra. 

The explicit time dependence appearing on the right-hand side of \eqref{flowrmch} can be understood by computing the time derivative of e.g. $\bar P^{KM}$, where as usual $\frac{d}{dt} = \p_t-\{H,\, \cdot \, \}$. One finds that $ \frac{d}{dt} \bar P^{KM}   = \{\Dl H, \bar P_{\bar \eta } \} \neq 0$ because, as a result of the first equation,  $ \Dl H= \om E_R/R_v $ .  

Given the above form of the flow equations, it is convenient to define 

\be
\Dl' = \Dl -  \frac{ w E_R t}{R_v}
\ee
which annihilates all of the (rescaled) conserved charges.

\subsection{Relating the two}

To summarize, we found that the quantum version of the rescaled conserved charges $Q_f, P_\eta, R_v \bar Q_{\bar f}$ and $\bar P_{\bar \eta}^{KM}$, which we will  collectively denote as $\L$, are  annihilated by  the operator \eqref{dldef} 

\be
\hat{\mathcal{D}}_\l \L = \p_\l  \L - i [ \O_{tot},  \L]= 0
\ee 
(or, $\hat{\mathcal{D}}_\l' \L =0$, if we work at $t \neq 0$). On the other hand, the states satisfy the flow equation \eqref{flowstates},
which involves an  additional diagonal  operator $D$, which can be rather complicated. 

We now consider two energy eigenstates, $|n_0\rangle$ and $|n_0'\rangle$,  that in the undeformed CFT are related by the action of a symmetry generator, $|n_0'\rangle= \L_{(\l=0)}|n_0\rangle$, which can be any of the Virasoro or Ka\v{c}-Moody generators. Our goal is to find a new operator, $\widetilde \L$,  that relates the corresponding flowed states in the deformed CFT, i.e.  $|n'_\l\rangle= \widetilde \L \, |n_\l\rangle$.  The flow equation \eqref{flowstates} for the states then implies that the flow equation for the corresponding operators is 

\be
\p_\l \widetilde \L - i [ \O_{tot} -D,\widetilde  \L] =0 \label{flowLwt}
\ee
The  solutions to the two flow equations are related by  $ \widetilde \L = e^X \L e^{-X}$ where $X$ must satisfy

\be
[\widetilde \L,  (\p_\l e^X - i[\O_{tot}, e^X])e^{-X}-D]  =0
\ee
for \emph{any} $\widetilde \L$. This implies that the second argument either vanishes, or it is proportional to the identity or some other operator that commutes with all the $\widetilde \L$. Assuming for simplicity that it vanishes, we can write

\be
D  = (\p_\l e^X -i [\O_{tot}, e^X])\,  e^{-X}
\ee 
Noting that 
\be
(\p_\l e^X) \, e^{-X} = \p_\l X + \frac{1}{2} [X, \p_\l X] + \frac{1}{3!} [X,[X,\p_\l X]] + \ldots
\ee

\be
e^X \O_{tot} e^{-X} = \O_{tot} +[X,\O_{tot}]+\frac{1}{2}[X,[X,\O_{tot}]] + \ldots
\ee
the above equation can be written as 

\be
D = \hat{\mathcal{D}}_\l X +  \frac{1}{2} [X, \hat{\mathcal{D}}_\l X] + \frac{1}{3!} [X,[X,\hat{\mathcal{D}}_\l X]] + \ldots \;, \;\;\;\;\;\; \hat{ \mathcal{D}}_\l X \equiv \p_\l X - i [\O_{tot}, X] \label{eqnX}
\ee
This result gives us a way to construct $X$, and therefore $\widetilde \L$, if we know $\L$ and $D$. If we work at $t\neq 0$, then $\Dl$ should be replaced by $\Dl'$, $\O_{tot}$ by $\O_{tot} (t) -w E_R t/R_v$ and $D$ by $D_{tot} (t) \equiv D(t)+ \p_\l E \, t$, as follows from \eqref{tdepflow}.

As we already explained, finding $D$ to all orders is a rather difficult task, but we can certainly attempt this exercise perturbatively.   Since at $\l =0$, $D = \D \O = (w \chi_0 - E_R \phi_0)/R$, where $\phi_0$ is known to implement spectral flow in a CFT, we will henceforth denote the $\widetilde \L$ as the ``spectrally flowed'' operators, in this case by an energy-dependent amount. This connection will be made significantly  more precise in the next section. 

\section{The spectrally flowed generators \label{flop}}

\subsection{Perturbative construction of the spectrally flowed generators}

In this section, we attempt to solve the equation \eqref{eqnX} for $X$ perturbatively, for $D$ given in \eqref{Dpert}, and use the solution to find the first few terms in the $\l$ expansion of the flowed generators. This can be done  by assuming $X$ has an expansion of the form

\be
X= \l \O_1 + \l^2 \O_2 +  \ldots  \;\;\;\; \Rightarrow \;\;\;\;\; \mathcal{D}_\l X = \O_1 + \l \mathcal{D}_\l \O_1 + 2 \l \O_2 + \l^2 \mathcal{D}_\l \O_2 +  \ldots \label{pertX}
\ee
where the $\O_n$ are in general non-linear functions of $\l$. We would moreover like to work at $t \neq 0$ so, according to our previous discussion,  $\mathcal{D}_\l$ should  be replaced by $\mathcal{D}'_\l$ and 
\be
D_{tot} (t) \equiv  D(t)+ \p_\l E\, t= D_0 - \l (\mathcal{D}'_\l D_0)_{no \; z.m.}+ \O(\l^2) + \frac{2 Q_K E_R}{R-\l Q_K}\, t 
 \ee
where $D_0 = \Delta \O-w E_R  t/R_v$. Since we know $D_{tot}$ to first order in $\l$, we can thus find $X$, and consequently $\widetilde \L$, to second order.  
%
%To recapitulate, to first order in $\l$  we   have
%\be
%\O_1 = D_0+ 2 Q_K E_R t\;, \;\;\;\;\;\;\;\O_2 = -  \mathcal{D}'_\l D_0+ \frac{1}{2} (\mathcal{D}'_\l  D_0)_{z.m.}+ Q_K^2 E_R t%\;, \;\;\;\;\;\;\;\O_3 = -\frac{1}{3} \mathcal{D}'_\l \O_2-\frac{1}{6} [\O_2,\O_1]+??
%\ee
%Thus, we should understand the flow equation that $\D \O$ itself satisfies
%
%where
Evaluating
\be
 \mathcal{D}'_\l D_0 = \frac{w}{R_v} (\D \O - w E_R\frac{ t}{R_v})-\frac{ w}{R} \int d\s   \frac{\H_R}{\sqrt{}} \frac{ wt+R \hat \phi(\s)}{R_v}  -\frac{ E_R}{R} \int d\s \frac{\J_-+\l \H_R/2}{\sqrt{}}  \frac{wt+ R \hat \phi(\s)}{R_v}
%& = & w \mathcal{D}_\l \chi_0  - E_R \mathcal{D}_\l \phi_0 = w \int d\s \left[ \frac{\H_R}{\sqrt{}}(w-\hat \phi(\s)) + w \hat A\right] - E_R \int d\s \frac{\J_-+\l \H_R/2}{\sqrt{}} \hat \phi - \frac{E_R w}{R} \{\chi_0,\phi_0\} \nonumber\\
%&=& -w \int d\s  \frac{\H_R}{\sqrt{}} \hat \phi(\s)  - E_R \int d\s \frac{\J_-+\l \H_R/2}{\sqrt{}} \hat \phi - \frac{E_R w}{R} \{\chi_0,\phi_0\} 
\label{DlD0}
\ee
 gives us 
\be
D(t) = \frac{ w \chi_0 - E_R \phi_0}{R} - w E_R \frac{t}{R_v}  + \frac{\l w}{R_v} \int d\s \frac{\H_R}{\sqrt{}} \left(\hat \phi -\frac{\phi_0}{R}\right) + \frac{\l E_R}{R_v} \int d\s \frac{\J_-+\l\H_R/2}{\sqrt{}} \left(\hat \phi - \frac{\phi_0}{R}\right)+ %\frac{2 Q_K E_R}{1-\l Q_K} t +
 \ldots 
\ee
%
%\be
%\O_2 = \frac{ w}{R_v} \int d\s   \frac{\H_R}{\sqrt{}}\left( \frac{w t}{2R} +  \hat \phi- \frac{\phi_0}{2 } \right)  + \frac{E_R}{R_v} \int d\s \frac{\J_-+\l \H_R/2}{\sqrt{}} \left( \frac{w t}{2R} + \hat \phi- \frac{\phi_0}{2 }\right)    + Q_K^2 E_R t - \frac{w}{2 R_v} (\D \O - w E_R t/R_v)
%\ee
It is extremely useful to note that to this order,  $D(t)$ can be written as 

\be
D(t) \approx  \frac{w \widetilde \chi_0 - E_R \widetilde \phi_0}{R} - w E_R  \frac{t}{R_v} - \frac{\l \phi_0 E_R Q_K}{R^2} %+ \frac{2 Q_K E_R}{1-\l Q_K} t
+ \O(\l^2) \label{Dpertfancy}
\ee
where  the ``improved'' zero modes $\widetilde \phi_0$ and $\widetilde \chi_0$ are defined as
 
 \be
\widetilde \phi_0 \equiv \phi_0 - \frac{\l R}{R_v} \int d\s (\J_- + \l \H_R/2) \hat \phi
\;, \;\;\;\;\;\;\;\;
\widetilde \chi_0 \equiv \chi_0 + \frac{\l R}{R_v} \int d\s \H_R \hat \phi
\ee
The usefulness of introducing these quantities stems from the extremely simple Poisson  brackets they satisfy, to \emph{all orders} in $\l$. The Poisson brackets of   $\widetilde \phi_0$ with the left-movers are
 
\be
\left\{  \widetilde \phi_0,K_U \right\} =  \frac{1}{2}\;, \;\;\;\;\;\;\;\;\left\{  \widetilde \phi_0,\H_L \right\} = K_U \label{commwtphilm}
\ee
and with the right-movers

\be
\{\widetilde \phi_0 ,\, \bar Q_{\bar f} \} = \frac{R}{R_v}
\bar P^{KM}_{\bar f}% + \l w \int d\s \bar f (v/R_v)\,  \frac{\J_- + \l \H_R/2}{\sqrt{}}=\bar P^{KM}_{\bar f} +\l w \{\phi_0,\, \bar Q_{\bar f} \}
\;, \;\;\;\;\;\; \{\widetilde \phi_0 ,\, \bar P^{KM}_{\bar \eta} \}=\frac{R}{2 R_v}\int d\s  \bar \eta (1-\l \phi') \label{commwtphi0}
\ee
which implies  that $\widetilde \phi_0$ is the corrected operator implementing spectral flow in the $J\bar T$ - deformed CFT. The field $\widetilde \chi_0$ commutes with  $K_U, \H_L, E_R, \bar J_0$ and, for our particular choice of the constant $a$ in the Poisson brackets, with everything else\footnote{For $a\neq -1$, its non-zero commutators are 
%so clearly, this  generates a right-moving spectral flow. We can also check that for the right-moving KM
%%
%\be
%\{ \phi_0 -\l w \phi_0- \l \int d\s (\J_- + \frac{\l \H_R}{2}) \hat \phi , \bar P^{KM}_{\bar f} \} =\ \frac{1}{2}\int d\s  \bar \eta (1-\l \phi') %+ \frac{\l w}{2} \int d\s \bar \eta \left(1+ \l\frac{\J_-+\l \H_R/2}{\sqrt{}}\right)=\ \frac{1}{2}\int d\s  \bar \eta (1-\l \phi') +\l w \{ \phi_0 , \bar P_{\bar \eta}^{KM}\}
%\ee
%
\be
\{ \widetilde \chi_0 , \H_R\} = (1+a) R \p_\s  \frac{\H_R}{\sqrt{}} \;, \;\;\;\;\;\;\;\{ \widetilde \chi_0 , \J_-\} = (1+a) R \p_\s  \frac{\J_-}{\sqrt{}} \;, \;\;\;\;\;\;\; \{ \widetilde \chi_0 , \phi\} = - (1+a) R \frac{\J_-+\l/2 \H_R}{\sqrt{}}
\ee
which implies% \emph{Re-check last!}
\be
\{\widetilde \chi_0 , \bar Q_{\bar f}\} = - \frac{R}{R_v} \bar Q_{\bar f'} \left(1+ a\right)
\;, \;\;\;\;\;\;\;\;
\{\widetilde \chi_0 , \bar P^{KM}_{\bar \eta}\} = - \frac{\bar P_{\bar \eta'}^{KM}R}{R_v} \left(1+a\right)\;,\;\;\;\;\;\;\; \{ \widetilde \phi_0, \widetilde \chi_0\}=   \frac{R^2\bar Q_{\bar K}}{R_v} (1+a)
\ee
}.

It is also interesting to check the flow equations that $\widetilde \phi_0$ and $\widetilde \chi_0$   satisfy\footnote{More generally,
\be
\mathcal{D}_\l \widetilde \chi_0 = \mathcal{D}'_\l \widetilde \chi_0 =  \frac{w}{R_v} \widetilde \chi_0 +\frac{ w R}{R_v} E_R  (1+a) 
\;, \;\;\;\;\;\;\;\;\;
\mathcal{D}_\l \widetilde \phi_0 = - \frac{R \bar Q_{\bar K}}{R_v} w (1+a) \;, \;\;\;\;\;\;\;\;\;
\mathcal{D}'_\l \widetilde \phi_0 = \frac{wt R}{R_v^2} \bar Q_{\bar K}
\ee}

\be
 \mathcal{D}'_\l \widetilde \chi_0 =  \frac{w}{R_v} \widetilde \chi_0
\;, \;\;\;\;\;\;\;\;\;
\mathcal{D}'_\l \widetilde \phi_0 = \frac{wt R}{R_v^2} \bar Q_{\bar K}
\ee
Given the rewriting \eqref{Dpertfancy} of $D (t)$, it is best to consider a different identification of what is considered `zeroth' versus `first' order in its expansion with respect to $\l$, namely 

\be
\widetilde D_0 =  \frac{w \widetilde \chi_0 - E_R \widetilde \phi_0}{R} - w E_R  \frac{t}{R_v} + \O(\l^2) \;, \;\;\;\;\;\;\; \widetilde D_1= - \frac{\phi_0 E_R Q_K}{R^2} + \O(\l) 
\ee 
Then, the  coefficients of the perturbative expansion \eqref{pertX} of $X$, obtained using \eqref{eqnX}, are given by 
%
%we can consider  an alternate split of $X$ into $\O_{1,2}$, which makes computations significantly easier. This is $X= \l \widetilde{\O}_1  +\l^2 \widetilde{\O}_2 $, with
%
\be
\O_1 = \frac{w \widetilde \chi_0 - E_R \widetilde \phi_0}{R} -w E_R \frac{t}{R_v} +\frac{2 Q_K E_R t}{R}\;, \;\;\;\;\;\; \mathcal{D}'_\l \O_1 =\frac{w}{R_v}  \frac{w \widetilde \chi_0 - E_R \widetilde \phi_0}{R} + \frac{2 Q_K  w E_R\, t}{R^2}  - \frac{wt E_R (Q_K + \om)}{R^2}
\ee
which implies that, to zeroth order in $\l$ 
\bea
R^2 \O_2 &=& - \frac{1}{2} \phi_0 E_R Q_K -  \frac{w(w \widetilde \chi_0 - E_R \widetilde \phi_0)}{2} - Q_K w E_R t + Q_K^2 E_R t + \frac{1}{2} wt E_R (Q_K + \om)\nonumber\\
&\approx& - \frac{1}{2} \phi_0 E_R \bar Q_{\bar K}- \frac{\om^2}{2} \chi_0 +  E_R Q_K \bar Q_{\bar K} t + \frac{1}{2} wt E_R (Q_K + \om)
\eea
The above expressions for $ \O_{1,2}$ give us the classical limit of the operator $X$ entering the similarity transformation, up to $\O(\l^3)$. 

We would now like to check  the effect of the similarity transformation on the various conserved charges.  To pass from the quantum commutators to Poisson brackets, we note that the operator $X$ should have a factor of $\hbar ^{-1}$ in front, which cancels against the $\hbar$ factors in the commutators to yield the classical result

\be
\widetilde \L = e^X \L e^{-X } \;\;\leftrightarrow \;\; \widetilde \L_{cls} = \L + \l \{  \O_1, \L\} + \l^2 \{  \O_1, \L\} + \frac{\l^2}{2} \{  \O_1, \{  \O_1, \L\}\} + \O(\l^3)
\ee
Let us first work out the effect of this transformation on  $K_U$. Using
\be
\{  \O_1, K_U\} = - \frac{E_R}{2R } \;, \;\;\;\;\;\;\; \{ \O_2,K_U\} = - \frac{E_R \bar Q_{\bar K}}{2 R^2}  \{ \phi_0, K_U\} = - \frac{E_R \bar Q_{\bar K}}{4 R^2}  \;, \;\;\;\;\;\;\;\{  \O_1,E_R\} = - \frac{E_R\bar Q_{\bar K}}{R}
 \ee
we can readily show that
\be
\boxed{\widetilde  K_U= K_U - \frac{\l E_R}{2R } + \O(\l^3) \label{flowedKU}}
\ee
Note in particular that the zero mode of  $\widetilde  K_U$ is just $J_0$. This  implies that the spectrally flowed generators will commute with $J_0$, since  the commutator $[J_0, e^X \bar Q_{\bar f} e^{-X}] = e^X [e^{-X} J_0 e^X, \bar Q_{\bar f}] e^{-X} =e^X [Q_K, \bar Q_{\bar f}] e^{-X} =0$.% For this to be true for any charge $\bar Q_{\bar f}$, we should  basically find that $e^{-X} J_0 e^X = K_U$.

 Next, we would like to check what happens to the left-moving energy current $\H_L$. We evaluate
 
 \be
\{  \O_1, \H_L\} = - \frac{E_R K_U}{R} \;, \;\;\;\;\;\;\; \{ \O_2,\H_L\} = - \frac{E_R \bar Q_{\bar K}}{2 R^2}  \{ \phi_0, \H_L\} = - \frac{E_R \bar Q_{\bar K}}{2 R^2}  K_U 
\ee

\be \{ \O_1,- E_R K_U\} =  \frac{E_R^2}{2R} +\frac{E_R K_U \bar Q_{\bar K}}{R}
 \ee
which in turn implies that  
\be
\boxed{\widetilde \H_L = \H_L - \frac{\l E_R K_U}{R} + \frac{\l^2 E_R^2}{4 R^2}  + \O(\l^3)} \label{flowedHL}
\ee 
Note that  the transformation of $K_U$ and that of $\H_L$ correspond precisely to a spectral flow transformation with parameter $\l E_R$. The  zero mode of $\widetilde \H_L$ equals the left-moving energy $E_L^{(0)}$ in the  undeformed CFT. 

Let us now turn to the right-movers, starting with the Ka\v{c}-Moody generators $ \bar P^{KM}_{\bar \eta}$ in \eqref{rmch}. Using \eqref{commwtphi0}, we compute 
\be
\{\O_1, \bar P^{KM}_{\bar \eta} \}= - \frac{E_R}{2 R_v} \int d\s \bar \eta (1-\l \phi') + \frac{1}{R_v} \left(\frac{\widetilde \phi_0 -2 Q_K t}{R} + \frac{wt}{R_v}\right) \bar P_{\bar \eta'}^{KM}
\ee
\be
\{ \O_2, \bar P^{KM}_{\bar \eta} \}\approx - \frac{E_R \bar Q_{\bar K}}{4 R_v R} \int d\s \bar \eta (1-\l \phi') + \frac{\bar Q_{\bar K}}{2R^2} (\widetilde \phi_0 -2 Q_K t ) \bar P_{\bar \eta'}^{KM}% +\frac{\om^2}{2} (1+a)\bar P_{\bar \eta'}^{KM}
 -  \frac{wt }{2R^3}  (Q_K + \om) \bar P_{\bar \eta'}^{KM}
\ee
It is useful to treat separately the case in which $\eta = I$ (the identity), for which

\be
\tilde{\bar P}_{I}^{KM}  = \left(\bar J_0 + \frac{\l E_R}{2}\right) - \frac{\l E_R}{2}- \frac{\l^2}{4 R} E_R \bar Q_{\bar K} + \frac{\l^2}{2} \{\O_1, - E_R/2\}= \bar J_0 +\O(\l^3)
\ee
as expected from spectral flow. For $\bar \eta \neq I$, we compute 
\be
R^2 \{ \O_1,\{ \O_1, \bar P^{KM}_{\bar \eta} \}\}= \left(\frac{\phi_0-2Q_K t}{R}+ \frac{wt}{R_v}\right) \bar Q_{\bar K} \bar P_{\bar \eta'}^{KM}  +\left(\frac{\phi_0-2Q_K t}{R}+ \frac{wt}{R_v}\right)^2  \bar P_{\bar \eta''}^{KM}
\ee
%In particular, all the dependence on the $a+1$ terms drops out. 
The transformed  $ \bar P^{KM}_{\bar \eta}$ is then, for $\bar \eta \neq I$
\bea
 \tilde{ \bar P}_{\bar \eta}^{KM}\!\!\!\! &=& \!\!\! \bar P_{\bar \eta}^{KM}\! + \frac{\l}{R_v} \!\left(\frac{\widetilde \phi_0 -2 Q_K t}{R}+ \frac{wt}{R_v}\right)  \!\bar P_{\bar \eta'}^{KM}\!+\frac{\l^2}{2R^2}\!\left(\frac{\widetilde \phi_0 -2 Q_K t}{R} + \frac{wt}{R_v}\right)^2 \!\!\!\! \bar P_{\bar \eta''}^{KM}\! +\frac{\l^2}{R^3} \!\left[   \bar Q_{\bar K}(\phi_0-2Q_K t)% + \om^2 (1+a)
- \om^2 t\right]\!\!\bar P_{\bar \eta'}^{KM}\! + \O(\l^3)
 \nonumber\\
&\approx & \!\!\! \bar P_{\bar \eta}^{KM} + \hat \l (1+\hat \l Q_K) \left(\frac{\widetilde \phi_0 - 2 Q_K t +\om t -\hat \l \om \bar Q_{\bar K} t}{R}\right)  \bar P_{\bar \eta'}^{KM} + \frac{\hat \l^2}{2} \left(\frac{\widetilde \phi_0 - 2 Q_K t+\om t}{R} \right)^2\!\! \bar P_{\bar \eta''}^{KM} + \O(\l^3) \label{pwtpert}
\eea
where we have introduced the dimensionless $J\bar T$ coupling $\hat \l=\l/R$.
It is not hard to check that this expression is conserved to the given order in $\l$, using 

\be
\{ H, \widetilde \phi_0\} = - Q_K - \frac{\bar Q_{\bar K} R}{R_v}
\ee
Using $\{ P, \widetilde \phi_0\} = - Q_K + \frac{\bar Q_{\bar K}R}{R_v} \approx - w (1-\hat \l \bar Q_{\bar K})$ and the approximate expansion above, we also find

\be
\{ P, \tilde{ \bar P}_{\bar \eta}^{KM} \}  = \frac{1}{R} \, \tilde{ \bar P}_{\bar \eta'}^{KM}  + \O(\l^3)
\ee
i.e., acting with the spectrally flowed right-moving generator preserves the quantization of the momentum.

Finally, moving on to the right-moving pseudoconformal generators, we find

\be
\{ \O_1, \bar Q_{\bar f} \}=   - \frac{E_R \bar P^{KM}_{\bar f}}{ R_v}  + \frac{1}{R_v} \left(\frac{\widetilde \phi_0 -2 Q_K t}{R}+ \frac{wt}{R_v} \right) \bar Q_{\bar f'}
\ee

\be
\{ \O_2, \bar Q_{\bar f} \}=- \frac{E_R \bar Q_{\bar K}}{2 R_v R} \bar P_{\bar f}^{KM} + \frac{\bar Q_{\bar K}}{2 R_v R^2} (\widetilde \phi_0 -2 Q_K t ) \bar Q_{\bar f'}% + \frac{\om^2}{2} (1+a) \bar Q_{\bar f'}
 - \frac{wt (Q_K+\om)}{2 R^3} \bar Q_{\bar f'}
\ee

\be
\{\O_1, \{ \O_1, \bar Q_{\bar f} \}\}= - \frac{E_R}{R_v} \{ \O_1, \bar P_{\bar f}^{KM}\} + \frac{E_R \bar Q_{\bar K}}{R_v R } \bar P_{\bar f}^{KM} + \frac{\frac{\phi_0-2 Q_K t}{R}+ \frac{wt}{R_v}}{R_v} (\bar Q_{\bar K} \bar Q_{\bar f'} -E_R  \bar P_{\bar f'}^{KM} )+\frac{(\frac{\phi_0-2 Q_K t}{R}+ \frac{wt}{R_v})^2}{R_v^2}\bar Q_{\bar f''}
\ee
Let us first check the case $f=const$. Remembering that the orginal right-moving pseudoconformal generator that satisfies $\Dl' \L=0$  is $\bar Q_{\bar f} R_v$, applying the similarity transformation to it yields 

\be
 \widetilde E_R \, R= %E_R  R_v - \l E_R \bar Q_{\bar K} - \frac{\l^2}{2} E_R   \bar Q_{\bar K}^2 + \frac{\l^2}{2} ( \frac{E_R^2}{2}+ \bar Q_{\bar K}^2 E_R) + \O(\l^3) = 
E_R R_v - \l E_R \bar J_0 - \frac{\l^2}{4} E_R^2 + \O(\l^3)
\ee
in perfect agreement with our expectation, $R E_R^{(0)}$. The factor of $R$ on the left-hand side has been included for dimensional reasons. For general $f $, we find
\bea
\widetilde{\bar Q}_{\bar f} \,  R&=& \bar Q_{\bar f} R_v - \l E_R \bar P^{KM}_{\bar f}  +\frac{\l^2 E_R^2}{4}  \d_{f=I} + \l \left(\frac{\widetilde \phi_0 - 2 Q_K t}{R}+ \frac{wt}{R_v} \right) \bar Q_{\bar f'}+\frac{\l^2}{2 R^2} (\phi_0-2Q_K t) (2 \bar Q_{\bar K} \bar Q_{\bar f'}- 2 E_R \bar P^{KM}_{\bar f'} ) \nonumber \\
&+&  \frac{\l^2}{2R} \left(\frac{\phi_0 -2 Q_K t}{R}+ \frac{wt}{R_v}\right)^2 \bar Q_{\bar f''}   - \frac{\l^2\om t}{2 R^2}  (Q_K + \om) \bar Q_{\bar f'} + \frac{\l^2 \om t}{2 R^2} (\bar Q_{\bar K} \bar Q_{\bar f'} - 2 E_R \bar P_{\bar f'}^{KM})% + \frac{\l^2 \om^2}{2} (1+a) \bar Q_{\bar f'}
\eea
As before, this can be organised as  the following perturbative expansion
\bea
\widetilde{\bar Q}_{\bar f}\, R& = & R_v \left(\bar Q_{\bar f} + \hat\l (1+\hat \l Q_K) \,\frac{\widetilde \phi_0-2 Q_K t + \om t (1-\hat \l \bar Q_{\bar K})}{R} \bar Q_{\bar f'} + \frac{\hat \l^2}{2R^2} (\phi_0-2 Q_K t + \om t)^2 \bar Q_{\bar f''}\right) \nonumber\\ && - \l E_R \left(\bar P_{\bar f}^{KM} + \frac{\l (\phi_0 -2 Q_K t + \om t)}{R^2}\bar P_{\bar f'}^{KM}\right) +\frac{\l^2 E_R^2}{4}  \d_{f=I} +\O(\l^3) \label{Qwtpert}
\eea
This has precisely the correct form to yield  a conserved charge and an integer-quantized momentum, as one can check by computing its Poisson brackets with $H$ and $P$.

\subsection{An all-orders proposal}

The result of the perturbative analysis we have just performed is that the symmetry generators that  act properly  on the eigenstates of the system are given by a kind of  energy-dependent spectral flow.  While the form of the resulting left-moving generators, $\widetilde K_U$ and $\widetilde \H_L$ in \eqref{flowedKU} and \eqref{flowedHL}, matches precisely to what we expect from spectral flow with parameter $\l E_R$, the form of the right-moving generators \eqref{pwtpert} and \eqref{Qwtpert} is significantly more involved. In particular, while it is nothing but natural that  \eqref{flowedKU}  and  \eqref{flowedHL} should represent the full expressions for the left-moving flowed generators to all orders in $\l$, it is also clear that the $\l$ expansion of the right-moving generators will  contain an infinite number of terms.

In this section, we will make a proposal for an all-orders (formal) expression for the right-moving generators, starting from the assumption that \eqref{flowedKU}  and  \eqref{flowedHL} are the correct expression for the flowed left-moving currents to all orders in $\l$. Our main tool will be the fact that the  charge algebra is preserved by the flow \eqref{flowLwt}, and therefore  the spectrally flowed left and right generators should commute to all orders in $\l$.

Our analysis will proceed in two steps. First, we will construct combinations of the right-moving conserved charges that commute with the left-moving spectrally flowed currents, and show that these building blocks satisfy the correct Poisson brackets with the energy, momentum and the global $U(1)$ charges  to have, upon quantization, a consistent action on the Hilbert space. Then, we find linear combinations of these blocks that satisfy the expected flow equation, with an operator $D$ we will similarly derive. 

Let us start by analysing the building block, $\widetilde {\bar \P}_{\bar \eta}$, for the  right-moving $U(1)$ generator. The requirement  that it commute with all the left-moving charges (or, alternatively, the currents), reads

\be
\{ K_U - \frac{\l}{2R} E_R , \widetilde {\bar \P}_{\bar \eta}\} =0 \;, \;\;\;\;\;\;\;\{  \H_L -\frac{\l E_R K_U}{R} + \frac{\l^2}{4R^2} E_R^2 ,  \widetilde {\bar \P}_{\bar \eta} \} =0 \label{commleft}
\ee
where  $\bar \P_{\bar \eta}$ can in fact be \emph{any} right-moving current. Note the second equation follows from the first if 

\be
\{ \H_L, \widetilde {\bar \P}_{\bar \eta}\} =2 K_U \{ K_U,\widetilde {\bar \P}_{\bar \eta}\} \label{relHLKUcomm}
\ee
which can \emph{a posteriori} be checked to be the case.  Remembering that

\be
\{ \widetilde \phi_0, K_U\} = \frac{1}{2} \;, \;\;\;\;\;\{ \widetilde \phi_0, E_R\} = \frac{\bar Q_{\bar K}R}{R_v} \;, \;\;\;\;\; \{ E_R , \bar P_{\bar \eta} \} = - \frac{1}{R_v} \bar P_{\bar \eta'}
% \{ \widetilde \phi_0, J_0\} =\{ \widetilde \phi_0, \bar J_0\} = \frac{1-\l Q_K%}{2 (1-\l w)}  
\ee
%
%\be
%\{ \widetilde \phi_0,\bar P_{\bar \eta}\} = \frac{1}{2(1-\l w)} \int d\s \bar \eta (1-\l\phi') \;, \;\;\;\;\;\{ \widetilde \phi_0,\bar Q_{\bar f}\} = \frac{1}{1-\l w} \bar P_{\bar f}
%\ee
 a natural Ansatz for $\widetilde {\bar \P}_{\bar \eta}$ (for $\eta \neq I$) is

\be
\widetilde {\bar \P}_{\bar \eta} = {\bar P}_{\bar \eta} + \hat \l  a_1 \widetilde \phi_0\,  \bar P_{\bar \eta'} + \frac{\hat \l^2  a_2}{2} \,  \widetilde \phi_0^2 \, \bar P_{\bar \eta''} + \ldots  \label{ansatzP}
\ee
%where the $\ldots$ stand for extra terms that commute with $K_U$ and $E_R$. In particular, this Ansatz refers specifically to $\eta \neq I$. 
Plugging into \eqref{commleft}, we find the recursion relation 

\be
a_{n+1} = \frac{a_n}{R-\l Q_K} \;, \;\;\;\;\; a_0=1
\ee
which implies that the solution is simply

\be
\widetilde {\bar \P}_{\bar \eta} = {\bar P}_{\bar \eta}+\frac{\hat \l  \widetilde \phi_0 }{R-\l Q_K}{\bar P}_{\bar \eta'}  +\frac{\hat \l^2  \widetilde \phi_0^2 }{2(R-\l Q_K)^2}{\bar P}_{\bar \eta'}   + \ldots \label{solPwt}
\ee
 Note that in the case of the Ka\v{c}-Moody current $P^{KM}_{\bar \eta}$, the first three terms agree precisely with the result \eqref{pwtpert} of the perturbative analysis  of the previous section, at $t=0$. One can also check that   \eqref{relHLKUcomm} holds, using \eqref{commwtphilm}, \eqref{commHLKUHR} and the commutators of $\phi$ with $\H_L$ and $K_U$. 

Note the above is a formal expression in that $\phi_0$, and thus $\widetilde \phi_0$, is not a well-defined operator. However,  its exponential is expected to be. Using the $\widetilde \phi_0$ Poisson brackets 

\be
%\{\widetilde  \phi_0, E_R  \} = \frac{\bar Q_{\bar K}}{R_v} \;, \;\;\;\;\;\;\;   
\{\widetilde \phi_0, J_0  \}=\{\widetilde \phi_0,\bar J_0  \}=R\, \frac{R-\l  Q_{ K}}{2 R_v} 
\ee
with $Q_K \equiv J_0 + \l E_R/2$ and $\bar Q_{\bar K} \equiv\bar J_0 +\l E_R/2 $ that
we presented earlier, we can easily check that the charges of this combination are as expected, namely%\emph{Fix factors of $R$!}

\be
\{J_0, \widetilde {\bar \P}_{\bar \eta} \} = \{\bar J_0, \widetilde {\bar \P}_{\bar \eta} \} = \frac{\l}{2 R_v} \widetilde {\bar \P}_{\bar \eta'} - \frac{ (R-\l Q_K)}{2 R_v} \frac{\l}{R-\l Q_K} \widetilde {\bar \P}_{\bar \eta'} =0 
\ee
The Poisson bracket with the right-moving energy is

\be
\{E_R, \widetilde {\bar \P}_{\bar \eta} \} = -  \frac{\l}{R-\l Q_K} \frac{\bar Q_{\bar K}}{R_v} \widetilde {\bar \P}_{\bar \eta'} - \frac{1}{R_v}  \widetilde {\bar \P}_{\bar \eta'}= - \frac{1}{R-\l Q_K}  \widetilde {\bar \P}_{\bar \eta'} \label{commwtPER}
\ee
while the commutator with $\H_L$  is given by 

\be
\{\H_L, \widetilde {\bar \P}_{\bar \eta} \} =\frac{ \l K_U}{R} \{E_R, \widetilde {\bar \P}_{\bar \eta} \} = - \frac{\l K_U}{R(R-\l Q_K)} \widetilde {\bar \P}_{\bar \eta'}\label{commwtPHL}
\ee
These together  imply that  the Poisson bracket with the total momentum is

\be
\{P,\widetilde {\bar \P}_{\bar \eta} \}= \{ E_L-E_R,\widetilde {\bar \P}_{\bar \eta} \} = -\frac{1}{R} \, \widetilde {\bar \P}_{\bar \eta'}
\ee
and thus the action of $\widetilde {\bar \P}_{\bar \eta}$ in the Fourier basis increases the momentum  by an integer amount in units of the radius, which is now consistent with semiclassical quantization. The total energy is given by 

 \be
 \{ E_L+E_R,\widetilde {\bar \P}_{\bar \eta} \} = - \frac{R+\l Q_K}{R-\l Q_K} \frac{ \widetilde {\bar \P}_{\bar \eta'}}{R}
 \ee
To ensure conservation of the charges, one %needs to modify their time dependence as 
%
%\be
%v \r v + \frac{t}{R_v} - \frac{1+\l Q_K}{1-\l Q_K} \frac{t}{R}= v + \frac{\l t}{R_v}  \frac{w-2 Q_K + \l w Q_K}{1-\l Q_K}
%\ee
%Conservation can be made manifest by using the block
should,  for $t\neq 0$, replace $\widetilde \phi_0 $ by the block

\be
\widetilde \phi_0 - \left(Q_K + \frac{\bar Q_{\bar K} R}{R-\l w}\right) t \label{consblock}
\ee
which is conserved by itself. This agrees precisely with what happened in our previous perturbative analysis, and  makes it manifest that each term in the sum \eqref{solPwt} is separately conserved. %, note that $\phi_0$ is not a good operator, reason for which we should exponentiate this. 

It is also interesting to note that \eqref{commwtPER} implies that the  spectrally flowed right-moving energy $E_R^{(0)}$ is also changed by an integer amount

\be
\left\{E_R^{(0)}, \widetilde {\bar \P}_{\bar \eta} \right\} =\left\{E_R- \frac{\l J_0 E_R}{R} - \frac{\l^2 E_R^2}{4 R^2} , \widetilde {\bar \P}_{\bar \eta} \right\} =\left\{E_R, \widetilde {\bar \P}_{\bar \eta} \right\} \left(1-\frac{\l Q_K}{R}\right) = -  \frac{1}{R} \, \widetilde {\bar \P}_{\bar \eta'}
\ee
as expected from the fact that it is the global mode $(\widetilde{\bar L}_0)$ of the spectrally flowed algebra.

 An identical analysis  for the case of the pseudoconformal generators shows that they must appear in the combination
\be
\widetilde{\bar \Q}_{\bar f} = \bar Q_{\bar f} + \frac{\hat \l \widetilde{\phi}_0}{R-\l Q_K} \bar Q_{\bar f'}   +\frac{\hat\l^2  \widetilde \phi_0^2 }{2(R-\l Q_K)^2}{\bar Q}_{\bar f''}   +  \ldots \label{solQwt}
\ee
and their commutation relations are exactly analogous with those of $ \widetilde {\bar \P}_{\bar \eta}$. 

As already explained, the formal expressions $\widetilde{\bar \P}_{\bar \eta}$ and $\widetilde{\bar \Q}_{\bar f}$ are not exactly the spectrally flowed $\widetilde \L$ generators, as the commutation requirement \eqref{commleft} is a weaker condition than the flow equation \eqref{flowLwt}. Instead, they represent the building blocks  of the  spectrally flowed generators. To find which linear combination of them represents the $\widetilde \L$, we now turn to the flow equation they satisfy. 

It turns out that  the flow operator $D$ can also be fixed by its commutation relations with $K_U$ and $\H_L$, upon choosing a judicious Ansatz. Assuming that the spectrally flowed left-moving generators are given precisely by \eqref{flowedKU} and \eqref{flowedHL}, the flow equation \eqref{flowLwt} they are expected to satisfy and the fact \eqref{flowleftcurr} that they are annihilated by $\Dl$ fix the commutation relations of $D$ to  to all orders to
\be
\{ D, K_U - \frac{\l}{2R} E_R\}=\mathcal{D}_\l (K_U - \frac{\l}{2R} E_R) = - \frac{1}{2 R_v} E_R \;,\;\;\;\;\;\;\;\{ D, \H_L -\frac{\l E_R K_U}{R} + \frac{\l^2 E_R^2}{4R^2} \} =- \frac{E_R}{R_v} (K_U - \frac{\l E_R}{2R})
\ee
%Since $D$ is diagonal in the energy eigenbasis, we expect $[D,E_R] =0$. This may be a bit hard to establish for the non-normalizable operators we consider. It can be easily seen that the above implies that $\{D,\H_L\} = 2 K_U \{ D, K_U\}$. We seem to have $\{D, E_R\} \neq 0$ for the perturbative solution at least.
%
We make the following Ansatz for $D$
\be
D = a(\l) \widetilde \phi_0 + b(\l) \widetilde \chi_0 + c(\l)
\ee
where the operators $a,b,c$ commute with both $K_U$ and $E_R$. The first equation implies that 

\be
a - \frac{\l}{R} \{ D, E_R\} = - \frac{E_R}{R_v}\;, \;\;\;\;\mbox{where} \;\;\{ D, E_R\}= \frac{a \bar Q_{\bar K} R}{R-\l w}
\ee
thus yielding
\be
a= - \frac{E_R}{R-\l Q_K} 
\ee
in perfect agreemnet with our perturbative solution \eqref{Dpertfancy}. While the coefficient $b$ is not fixed,  this is not very important, since with our choice of  $\chi_0$ Poisson brackets, $\widetilde \chi_0$ commutes with all operators. We will set $b=w/R$ to match with the perturbative answer, and $c=0$, at least on the $t=0$ slice.

We would now like to show that 

\be\boxed{
\widetilde {\bar P}^{ KM}_{\bar \eta}\equiv \widetilde{\bar \P}_{\bar \eta} + \frac{\l}{2} \widetilde{\bar \Q}_{\bar \eta} - \frac{\l E_R}{2}\,\d_{\bar \eta =I}} \label{flowedPbar}
\ee
and
\be\boxed{
R\,\widetilde{\bar Q}_{\bar f} \equiv R_v  \widetilde{\bar \Q}_{\bar f} - \l E_R \widetilde{\bar \P}^{KM}_{\bar f}\!\! - \frac{\l^2 E_R^2}{4} \, \d_{\bar f=I}} \label{flowedQbar}
\ee
precisely satisfy the flow equation \eqref{flowLwt} for the above choice of $D$. We compute, for $\eta \neq I$ and  at $t=0$
\be
\mathcal{D}_\l \widetilde{\bar P}^{KM}_{\bar \eta}   = \mathcal{D}_\l  \left(\frac{\hat \l \widetilde \phi_0}{R-\l Q_K} \right)\widetilde{\bar P}^{KM}_{\bar \eta'}  = \frac{\widetilde \phi_0}{(R-\l Q_K)^2} \widetilde{\bar P}^{KM}_{\bar \eta'}=\left\{ \frac{ w \widetilde \chi_0}{R} - \frac{E_R \widetilde \phi_0}{R-\l Q_K}, \widetilde{\bar P}^{KM}_{\bar \eta}\right\}
\ee
%
%\be
%\mathcal{D}'_\l \widetilde{\bar P}^{KM}_{\bar \eta}   = \mathcal{D}'_\l  \left(\frac{\l (\widetilde \phi_0- Q_K t - \bar Q_{\bar K} t/R_v )}{1-\l Q_K} \right)\widetilde{\bar P}^{KM}_{\bar \eta'}  = \frac{\widetilde \phi_0- Q_K t - \bar Q_{\bar K} t/R_v}{(1-\l Q_K)^2} \widetilde{\bar P}^{KM}_{\bar \eta'}
%\ee
%Noting that\footnote{For reference, $\{\widetilde \phi_0,\frac{\l (\widetilde \phi_0- Q_K t - \bar Q_{\bar K} t/R_v )}{1-\l Q_K} \} = - \frac{\l t}{(1-\l Q_K)^2} $.}
%
%\be
%\{ \widetilde \chi_0 , \widetilde{\bar P}^{KM}_{\bar \eta} \} = - \frac{1+a}{1-\l Q_K}\widetilde{\bar P}^{KM}_{\bar \eta'} = (1+a) \{ E_R,  \widetilde{\bar P}^{KM}_{\bar \eta'} \}
%\ee
%which can be shown to coincide with
%%
%\be
%\mathcal{D}_\l \widetilde{\bar P}^{KM}_{\bar \eta} = \{  w \widetilde \chi_0 - \frac{E_R \widetilde \phi_0}{1-\l Q_K}, \widetilde{\bar P}^{KM}_{\bar \eta}\}
%\ee
%Note the $\widetilde\phi_0$ commutators do not contribute because they are the same in $D$ as in $\widetilde{\bar P}$. Thus,  we should have 
%\be
%D= w \widetilde \chi_0 - \frac{E_R \widetilde \phi_0}{1-\l Q_K} -w (1+a)E_R
%\ee
%in perfect agreement with our perturbative example. 
The case $\eta = I$, for which $\widetilde{\bar P}^{KM}_{\bar I}\! \!\!=\bar J_0$ needs to be worked out separately, and it is easy to check that  %$\bar J_0$ satisfies

\be
\Dl \bar J_0 = \{ D, \bar J_0\}
\ee
which justifies  the  constant shift by $- \l E_R/2$. As far as the pseudoconformal charges are concerned, we find that, for $\bar \eta \neq I$

\be
\mathcal{D}'_\l \widetilde{\bar \Q}_{\bar \eta} - \left\{  \frac{w \widetilde \chi_0}{R} - \frac{E_R \widetilde \phi_0}{R-\l Q_K}, \widetilde{\bar \Q}_{\bar \eta}\right\} = \frac{\om}{R_v} \widetilde{\bar \Q}_{\bar \eta} +  \frac{R\, E_R }{R_v(R-\l Q_K)} \widetilde{\bar \P}^{KM}_{\bar \eta}
\ee
We can now easily show that the linear combination \eqref{flowedQbar} %$a(\l) \widetilde{\bar Q}_{\bar \eta}  + b(\l) \widetilde{\bar P}^{KM}_{\bar \eta}$, which
is exactly annihilated by $ \mathcal{D}_\l -D$. %We find
%
%\be
%\tilde{\mathcal{D}}_\l [a(\l) \widetilde{\bar Q}_{\bar \eta}  + b(\l) \widetilde{\bar P}^{KM}_{\bar \eta}] = \left(\tilde{\mathcal{D}}_\l a + \frac{w a}{R_v}\right) \widetilde{\bar Q}_{\bar \eta}+\left(\tilde{\mathcal{D}}_\l b + \frac{E_R a}{R_v (1-\l Q_K)}\right) \widetilde{\bar P}^{KM}_{\bar \eta}
%\ee
% with solution $a= R_v$ and $b= - \l E_R$. \emph{Does this hold only for $\eta \neq I$?} 
%For $\eta = I$, we should add $\frac{\l^2}{4} E_R^2$ in order to have the flow equation satisfied, which is \emph{exactly} what we expected.
The constant shift follows from the fact that  $\{\Dl - D, E_R\} = E_R Q_K/(R-\l Q_K) $, which implies in particular that 

\be
\left\{\Dl - D, E_R R_v - \l E_R \bar J_0 - \frac{\l^2}{4} E_R^2\right\} =0
\ee
The equations %\eqref{flowedKU}, \eqref{flowedHL}, 
\eqref{flowedPbar} and \eqref{flowedQbar} encode  the effect of the spectral flow by  an amount proportional to the right-moving energy on the right-moving charges. Due to the flow equation they obey, the spectrally flowed generators satisfy the same algebra as at $\l=0$, i.e. two commuting copies of the Witt-Ka\v{c}-Moody algebra.  The charges, defined at $t\neq 0$ via the replacement \eqref{consblock}, are conserved. The generators   of the global part of the algebra  are given by 

\be
\widetilde{ Q}_{f=I} = E_L^{(0)}\;, \;\;\;\;\;\; \widetilde{\bar Q}_{\bar f= I} = E_R^{(0)}%\;, \;\;\;\;\;\;\;\widetilde{ P}_I^{KM} = J_0\;, \;\;\;\;\;\;\;\;\widetilde{\bar P}_{\bar I}^{KM} = \bar J_0
\ee
whose sum \emph{does not} correspond to the energy  operator; rather, it is non-linearly related to it via \eqref{specasspfl}. This implies that while the Fourier modes of $\widetilde{\bar{Q}}$  and $\widetilde{\bar{P} }$ are integer-spaced with respect to $E_L^{(0)}$ and $E_R^{(0)}$ (as follows from the algebra), they have 
nontrivial commutation relations with $E_L$ and $E_R$, of the form \eqref{commwtPER}, \eqref{commwtPHL}.  This resolves the tension mentioned at the beginning of this article, between the integer spacing of the Virasoro descendants and the non-linear dependence \eqref{defeng} of the $J\bar T$ - deformed energy spectrum on the initial energy.

\section{Discussion}

In this article,  we have constructed a new, infinite set of symmetry generators in $J\bar T$ - deformed CFTs. Compared to the conserved charges  found in \cite{Guica:2020uhm}, these generators still  form two commuting copies of the Witt-Ka\v{c}-Moody algebra, but now their Poisson brackets with the global $U(1)$ charge and the momentum are  consistent with semiclassical quantization. It is therefore expected that in the quantum $J\bar T$ - deformed CFT on a cylinder, these generators will be acting properly on the Hilbert space of the theory, and thus the spectrum will organise  into Virasoro - Ka\v{c}-Moody representations.% They  do act properly on the Hilbert space, are conserved, and their  $L_0$ eigenvalues are simply those of the undeformed CFT.

The new symmetry generators are related to those of \cite{Guica:2020uhm} by a  type of energy-dependent spectral flow transformation, whose action on the various generators is given in \eqref{flowedKU}, \eqref{flowedHL}, 
\eqref{flowedPbar} and \eqref{flowedQbar}. The expressions \eqref{solPwt} and \eqref{solQwt} for the building blocks are somewhat formal and should be resummed in order to make sense. While the descendants obtained via the action of these symmetry generators will have integer-spaced spectrally flowed energies $E^{(0)}_{L,R}$, there is no tension with the known formula \eqref{defeng} for the deformed energies, as the latter are measured with respect to the non-flowed generators.  

As is clear from \eqref{defeng}, at large initial right-moving energy, the finite-size spectrum of $J\bar T$ - deformed CFTs will become complex. While this is only a problem for the theory on the cylinder, and not on the plane \cite{Guica:2019vnb}, it is interesting to remark that even on the cylinder, this problem is invisible at the level of the algebra and the spectrum of the flowed generators, which simply coincide with those of the undeformed CFT.  Rather, the complex energies arise solely as a result of the expression  \eqref{defHR} for the Hamiltonian, and  signal the breakdown of this square root formula. This  suggests that  the states of the deformed CFT on a cylinder may be well-behaved, and the only problem is with the operator  we use to measure the energy (which, however, is also the one used to evolve the system). 

There are several interesting  future directions. First, one should be able to pass from the classical construction of the symmetry generators to the quantum one by including the appropriate normal ordering prescriptions and check, at least perturbatively, whether their quantum commutators  work out as expected. A particularly interesting issue is that of the central extension, given that the classical Witt algebra is naturally expected to become Virasoro at the quantum level. The  holographic analysis of \cite{Bzowski:2018pcy} indicates a field-dependent central extension of the form $c (R/R_v)^2$ for the  generators defined in \cite{Guica:2020uhm}, where $c$ is the central charge of the original CFT. Given that the spectrally flowed generators are rescaled by a factor of $R_v$ with respect to the original ones suggests that the central extension in the flowed sector has a chance of being identical to that of the original CFT, in agreement with the fact that the $J\bar T$  deformation does not change the number of states. It would  be very interesting verify this, at least perturbatively. 

Other relevant questions are to understand the physical meaning of the spectrally flowed generators, and whether there is a geometric symmetry that they implement. It would also be interesting to see whether the new symmetry generators may have  applications also in the deformed CFT on the plane, e.g. in the construction of more complicated observables such as  correlation functions. Finally, it would be interesting to understand the role and physical  interpretation of the improved zero mode $\widetilde \chi_0$, whose action on the Hilbert space is surprisingly minimal. 

%
%
% \be
%\int d\tilde \s f(\s) g(\tilde \s) \p_\s \d(\s-\tilde \s) = f(\s) g'(\s) 
% \ee

\subsubsection*{Acknowledgements}

\noindent MG would like to thank Ruben Monten for collaboration on related subjects. This research was supported in part by  the ERC Starting Grant 679278 Emergent-BH and the Swedish Research Council grant number 2015-05333.

\appendix

\section{Summary of Poisson brackets in $J\bar T$ - deformed CFTs \label{pbapp}}

In this appendix, we collect the results of \cite{Guica:2020uhm} on the Poisson brackets of the various currents in the $J\bar T$ - deformed CFTs. These were computed using the expression they derived by solving the flow equation for the deformed right-moving Hamiltonian density $\H_R$

\be
\H_R = \frac{2}{\l^2} \left(1-\l \J_+ - \sqrt{(1-\l \J_+)^2 - \l^2 \H_R^{(0)}}\right) \label{defHR}
\ee
in terms of the undeformed one, $\H_R^{(0)}$, together with the commutation relations of the undeformed currents, $\H_R^{(0)}, \J_\pm$ and $\P$. The currents $\J_\pm $ represent the time components of the linear combinations $(J_\a\pm \tilde J_\a)/2$, where $J$ is represented as a $U(1)$ shift current for a scalar field $\phi$, and $\tilde J = \star d \phi$ is the corresponding topologically conserved current.  In Hamiltonian language

\be
\J_\pm = \frac{\pi \pm \phi'}{2}
\ee 
where $\pi$ is the momentum conjugate to $\phi$. Note the expression for $\H_R$ is symmetric under $\pi \leftrightarrow \phi'$, since it only depends on $\J_+$. From this,  we conclude that the $J\bar T$ and $\tilde J \bar T$ deformations, which differ by precisely this exchange, lead to the same deformed theory, at least at the classical level. 

The commutation relations derived in \cite{Guica:2020uhm} are
\be
\{\H_R(\s), \H_R(\tilde \s)\} %&=&\frac{- \H_R^{(0)}(\s) - \H_R^{(0)}( \tilde\s)+ \frac{\l^2}{2} \H_R (\s)  \H_R ( \tilde\s)}{\sqrt{\left(1-\l \J_+(\s)\right)^2-\l^2 \H_R^{(0)}(\s) } \sqrt{\left(1-\l  \J_+( \tilde\s)\right)^2-\l^2  \H_R^{(0)}( \tilde\s)}} \p_\s \d(\s-\tilde \s) \\
= - \left( \frac{\H_R(\s)}{\sqrt{\left(1-\l \J_+(\s)\right)^2-\l^2 \H_R^{(0)}(\s) } } + \frac{\H_R(\ts)}{\sqrt{\left(1-\l \J_+(\ts)\right)^2-\l^2 \H_R^{(0)}(\ts) } } \right) \p_\s \d(\s-\tilde \s) 
\ee
%\emph{Check!}
\be
 \{\P(\s), \H_R(\tilde \s)\} =% & \frac{  \H_R^{(0)} (\s)+\H_R^{(0)}(\tilde \s)+\l \J_+ ( \s)  \H_R (\tilde \s)}{  \sqrt{(1-\l  \J_+ (\tilde \s))^2-\l^2  \H_R^{(0)}(\tilde \s)}} \p_\s \d(\s-\tilde \s) \nonumber\\
% &=& 
 \left(\H_R(\s) + \frac{\H_R(\ts)}{  \sqrt{(1-\l  \J_+ (\tilde \s))^2-\l^2  \H_R^{(0)}(\tilde \s)}} \right)\p_\s \d(\s-\tilde \s)
 \ee

\be\{\H_R(\s), \J_+(\tilde \s)\} = \frac{\l \H_R(\s)}{2\sqrt{\left(1-\l  \J_+(\s)\right)^2-\l^2  \H_R^{(0)}}} \p_\s \d (\s-\tilde\s)
\ee

\be\{\H_R(\s), \J_-(\tilde \s)\} =-\frac{\J_- (\s)}{\sqrt{\left(1-\l  \J_+(\s)\right)^2-\l^2  \H_R^{(0)}(\s)}} \p_\s \d (\s-\tilde\s)
\ee 
Using the fact that $\J_+-\J_- = \phi'$ and that the right-hand side of the last two commutators are total $\tilde \s $ derivatives, we can deduce the commutator of $\H_R$ with $\phi$

\be
\{\H_R(\s),\phi(\tilde \s)\} = \frac{-\J_- (\s) - \l \H_R (\s)/2}{\sqrt{(1-\l \J_+(\s))^2-\l^2  \H_R^{(0)}(\s)}}\d (\s-\tilde\s)
\ee
Since this commutator is obtained by integration, in principle we could add an integration function of $\s$ to the right-hand side; however, such an addition would be quite unnatural, given that the commutator is local (i.e., proportional to a $\d$ function). 

The commutators of the momentum and the currents are the same as in the undeformed CFT

\be
\{ \P (\s),  \P (\tilde \s) \}= \left(\P(\s)+ \P (\tilde\s)\right) \p_\s \d(\s-\tilde \s)\;, \;\;\;\;\;\;\;\;\;
\{ \J_\pm (\s), \J_\pm (\tilde \s)\} = \pm \frac{1}{2} \p_\s \d (\s-\tilde \s)
\ee
%The commutators with the currents are

\be
\{ \P (\s),  \J_+ (\tilde\s)\}= \J_+  (\s)\p_\s \d (\s-\tilde\s) \;,\;\;\;\; % \{\J_+, \tilde \P\}  = \tilde J_+ \p_\s \d 
%\{ \H_R^{(0)}(\s),  \J_- (\tilde \s)\} = - \J_- (\s) \p_\s \d(\s-\tilde \s)
\ee
From here, one can deduce that 

\be
\{ \P (\s), \phi(\ts) \} = - \phi' (\s) \d(\s-\ts) \;, \;\;\;\;\;\;\;\; \{ \J_\pm (\s), \phi(\ts)\} = - \frac{1}{2} \d(\s-\ts)
\ee
We note that the zero modes $J_0 = \int d\s \J_+$,  $\bar J_0 = \int d\s \J_-$  commute with all the other currents in the theory, and their only non-zero commutator is with $\phi$. The winding  charge $w = J_0 -\bar J_0$ also commutes with $\phi$. This implies in particular that the field-dependent radius $R_v$ commutes with all the operators we consider.

Finally, one can work out the commutators of the chiral current $K_U = \J_+ + \frac{\l}{2} \H_R$ and of the left-moving Hamiltonian $\H_L = \H_R + \P$,  which take the very simple form

\be
\{ K_U (\s), K_U (\ts) \} = \frac{1}{2} \p_\s \d (\s-\ts)\;, \;\;\;\;\;\;\{ \H_L (\s), K_U (\ts) \} = K_U (\s) \p_\s \d(\s-\ts)
\ee

\be
\{ \H_L (\s), \H_L(\tilde \s)\} = (\H_L (\s) + \H_L (\tilde \s)) \p_\s \d(\s-\tilde \s) 
\ee
and their commutators with $\H_R$ are
\be
\{ \H_R(\s), K_U (\ts)\} = - \frac{\l \tilde \H_R}{2\sqrt{\tilde{}}} \d' \;, \;\;\;\;\;\; \{ \H_R(\s), \H_L (\ts)\} = \left( \tilde \H_R - \frac{\tilde \H_R}{\sqrt{\tilde{}}} \right) \d' \label{commHLKUHR}
\ee
which are total $\s$ derivatives. In particular, this implies that $\{ E_R, K_U\} = \{E_R, \H_L\} =0$.

 \section{Poisson brackets of the non-local field $\chi$ \label{pbchi}}
 
In this appendix, we derive the commutators of the  non-local field $\chi$, defined through $\p_\s \chi =\H_R $, with the various other fields in the theory.  

The Poisson brackets of $\chi$ are obtained by integrating the corresponding commutators of $\H_R$. Two of these Poisson brackets, namely

\be
\{ \chi(\s), K_U (\ts)\} = - \frac{\l \tilde \H_R}{2\sqrt{\tilde{}}} \d \;, \;\;\;\;\;\; \{ \chi(\s), \H_L (\ts)\} = \left( \tilde \H_R - \frac{\tilde \H_R}{\sqrt{\tilde{}}} \right) \d
\ee
are local, being proportional to $\d$ functions, and thus we do not include an integration function. Other commutators, however, are significantly more involved, and require working out the consistency conditions imposed by the various Jacobi identities that they satisfy.  

In the following, we will frequently use the notation $\sqrt{}= \sqrt{(1-\l \J_+(\s))^2 - \l^2 \H_R^{(0)}(\s)}$ and $\sqrt{\tilde{}}= \sqrt{(1-\l \J_+(\ts))^2 - \l^2 \H_R^{(0)}(\ts)}$, as well as the abbreviations $\tilde A = A(\ts)$ and $\d'=\p_\s \d(\s-\ts)$.

\subsection{Poisson bracket of $\chi$ with  $\H_R$}

The $\{\chi, \tilde{\H}_R\}$ Poisson bracket is given by integrating the $\{ \H_R, \tilde \H_R\}$ commutator

\be
\{ \chi(\s), \H_R(\ts)\} = - \frac{2 \H_R}{\sqrt{}} \d(\s-\ts) + \p_{\ts} \frac{\tilde{\H}_R}{\sqrt{\tilde{} }} \, \Theta(\s-\ts) + A(\ts) \label{defA}
\ee
where $A(\ts)$ is an integration function. This function needs to have winding, in order to cancel the dependence on the starting point in

\be
\{ \chi(\s), H_R\} = - \frac{ \H_R}{\sqrt{}} - \frac{ \H_R (0)}{\sqrt{(0)}} + \int_0^R d\ts A(\ts) \equiv - \frac{ \H_R}{\sqrt{}} + \int_0^R d\ts A_p(\ts)
\ee
where $E_R = \int \H_R d\s$  is the total right-moving energy. This equation  
 defines $A_p$, the periodic part of $A$, with the winding subtracted. To fix this function, we need to analyse the constraints coming from the various Jacobi identities that the Poisson bracket \eqref{defA} satisfies.

\subsubsection*{Constraints from time evolution}

A first constraint on $A$ comes from  analysing  the time dependence of the above commutator

\be
\frac{d}{dt} \{ \chi(\s), \H_R(\ts)\}= \p_t A (\ts) - \{H,\{ \chi(\s), \H_R(\ts)\}\} = - \{\{ H, \chi\},\tilde \H_R\}-\{ \chi,\{ H,\tilde \H_R\}\}
\ee
where $H = E_L + E_R$ is the total  Hamiltonian. Making use of 
\be
\{ H, \H_R\} = \p_\s (2 \H_R/\sqrt{} -\H_R)\;, \;\;\;\;\;\; \{ H, \J_+\} = - \p_\s (\J_+ + \l \H_R/\sqrt{})\;, \;\;\;\;\; \{ H, \H_R/\sqrt{} \} = \frac{1+\l K_U}{1-\l K_U} \p_\s (\H_R/\sqrt{})
\ee
It can be easily shown that the terms proportional to $\Theta, \d'$ and $\d$ functions cancel, and the constraint that we obtain on the function $A$ is 

\be
\p_t A - \{ H, A\} = \int_0^R d\ts \{ A_p (\ts) , \H_R\} - \p_\s \left(\frac{1+\l K_U}{1-\l K_U} A \right) \label{conseqA}
\ee
where $A_p$ represents the part of $A$ without the winding contribution. If we 
choose $A$ such that this term is absent, we find several qualitatively different solutions to the remaining equation. For example
 
 \be
R A \equiv \p_\s  \left(\s\frac{\H_R}{\sqrt{}} \right) + R \p_\s \hat A = \s \p_\s \frac{\H_R}{\sqrt{}}+\frac{\H_R}{\sqrt{}} + R a\,  \p_\s \frac{\H_R}{\sqrt{}} \label{fansatz}
\ee
solves the equations for an arbitrary constant $a$. Another solution  can be obtained by noting that   
 the field-dependent coordinate $v = \s -t - \l \phi$ satisfies 
\be
\frac{dv}{dt} = -1 - \{ H, v\} = - \frac{1+\l K_U}{1-\l K_U} v' 
\ee
and thus a  general solution to the equation with $A_p =0$ and the correct winding is

\be
A = \p_\s \left( \frac{v}{R_v} \frac{\H_R}{\sqrt{}} \right) + a\,  \p_\s \frac{\H_R}{\sqrt{}}
\ee
where the total derivative is necessary in order to ensure that $A_p=0$ above.

\subsubsection*{Constraints from the Jacobi identity with $K_U$}

To distinguish between the two solutions, we can check the Jacobi identity for $\{ K_U,\{ \chi_0, \tilde \H_R\} \}$. Using the expression  for the 
$\{\chi, \tilde K_U\}$ commutator, 
we obtain
\be
 \{ K_U, (R-\ts) \p_{\ts} \frac{\tilde{\H}_R}{\sqrt{\tilde{}}}  + R \tilde A\} + \frac{\l \d'}{2 \sqrt{}} \left[(R-\s ) \p_\s \frac{\H_R}{\sqrt{}} + R A \right] = \frac{\l}{2\sqrt{}} \p_\s \frac{\H_R}{\sqrt{}} \d(\s-\ts) 
\ee
Taking into account the fact that\footnote{This is most simply computed by using $\sqrt{} = 1-\l K_U$ and  the equivalence of distributions spelled out in footnote \ref{equivfoot}.} 
\be
\{ K_U, \frac{\tilde \H_R}{\sqrt{\tilde{}}} \} = \frac{\l}{2\sqrt{}} \p_\s \frac{\H_R}{\sqrt{}} \d(\s-\ts) \label{commKUHRsq}
\ee
we can reduce this to the following simple constraint on $A$

\be
\{ K_U, \tilde A\} + \frac{\l}{2\sqrt{}} A \d'=0 \label{constrKUa}
\ee
%We immediately see that the first solution will not work, for any $a$. , by checking
%
%\be
%(R-\s ) \p_\s \frac{\H_R}{\sqrt{}} + R A = \frac{\H_R}{\sqrt{}} + R (a+1) \p_\s \frac{\H_R}{\sqrt{}}
%\ee
%since the last term cancels from the LHS by itself, whereas the first  term cancels between the first term and the RHS, but nothing cancels the middle term on the RHS.
% Let us then try the solution with a $v$-type winding term. For this solution, we have
%\be
%(R-\s ) \p_\s \frac{\H_R}{\sqrt{}} + R A = R \p_\s \frac{\H_R}{\sqrt{}} (1-\frac{t}{R_v}) + \frac{R}{R_v} \left( \frac{\H_R}{\sqrt{}} (1-\l \phi') - \l \hat \phi \p_\s  \frac{\H_R}{\sqrt{}}  \right) = R \p_\s \frac{\H_R}{\sqrt{}} (1-\frac{t}{R_v}) - \frac{\l R}{R_v} \p_\s \left(\hat \phi \frac{\H_R}{\sqrt{}} \right)+ \frac{ \H_R}{ \sqrt{}}
%\ee
%As discussed, the first term just drops out of our consistency relation. For the other term, we compute
It is easy to check, using \eqref{commKUHRsq}, that the first Ansatz, \eqref{fansatz}, does \emph{not} satisfy the consistency requirement \eqref{constrKUa}. On the other hand, using the  commutator

\be
\{ K_U, \tilde v \frac{\tilde \H_R}{\sqrt{\tilde{}}} \} = - \l \frac{\tilde \H_R}{\sqrt{\tilde{}}}  \{ K_U, \tilde \phi\} + \tilde v \{ K_U, \frac{\tilde \H_R}{\sqrt{\tilde{}}} \} = \frac{\l}{2 \sqrt{}} \p_\s (v \H_R/\sqrt{}) \d
\ee
we can show that the second Ansatz does identically satisfy the consistency condition\footnote{
Making use (or not) of the relation

\be
\{ K_U, \tilde{\hat{\phi}}\frac{\tilde \H_R}{\sqrt{\tilde{}}} \} = \frac{\l \d(\s-\ts)}{2 \sqrt{}} \p_\s \left(\hat \phi \frac{\H_R}{\sqrt{}} \right) - \frac{R_v \H_R \d}{2 R (\sqrt{})^2}
\ee}. One can also check that the term $a \p_\s \H_R/\sqrt{}$ also satisfies it, so this does not fix $a$. 

\subsubsection*{Constraints from the Jacobi identity with $\H_R$}

We now look at the $\{\chi_0, \H_R\}$  Jacobi identity

\be
\{ \H_R, \{ \chi_0 , \tilde \H_R\}\}- \{ \tilde  \H_R, \{ \chi_0 , \H_R\}\} + \{ \chi_0, \{ \tilde \H_R, \H_R\}\} =0
\ee
We will also need
\be
\{ \H_R, \frac{\tilde \H_R}{\sqrt{\tilde{}}} \} = - \frac{2 \H_R^{(0)}}{\sqrt{}^3} \d'(\s-\ts) + \left(  \frac{1}{\sqrt{}} \p_\s \frac{\H_R}{\sqrt{}} -2\, \p_\s \frac{\H_R^{(0)}}{\sqrt{}^3}\right) \d (\s-\ts)
\ee
where, from \eqref{defHR},  $\H_R^{(0)} = \H_R (1-\l \J_+ - \frac{\l^2}{4} \H_R)$. Plugging in the expression for 

\be
\{\chi_0, \H_R \}= - 2 \H_R/\sqrt{} + (R-\s) \p_\s (\H_R/\sqrt{}) + R A
\ee
 we find the intermediate equation
\bea
&&R \left( \{ \H_R, \tilde A\} - \{ \tilde \H_R, A\} + \frac{A \d'}{\sqrt{}} + \frac{\tilde A \d'}{\sqrt{\widetilde{}}} \right) + (R-\ts) \{ \H_R, \p_{\ts} \frac{\tilde \H_R}{\sqrt{\tilde{}}} \}- (R-\s) \{ \tilde \H_R, \p_\s \frac{\H_R}{\sqrt{}} \} + \nonumber\\
&& \hspace{2cm}+ \left( \frac{R-\s}{\sqrt{}}\p_\s \frac{\H_R}{\sqrt{}} + \frac{2 \H_R^{(0)}}{\sqrt{}^3} + (\s \r \ts) \right) \d' =0
\eea
The difference
\be
(R-\ts) \{ \H_R, \p_{\ts} \frac{\tilde \H_R}{\sqrt{\tilde{}}} \}- (R-\s) \{ \tilde \H_R, \p_\s \frac{\H_R}{\sqrt{}} \}
\ee
can be manipulated using the criteria for when two distributions of the form 
 $E(\s,\ts) \d'' + F(\s,\ts) \d' + G(\s) \d $ are equivalent\footnote{These crieria are obtained by integrating against a test function $g (\ts)$, with the result
\be
g'' E + g' (2 \p_{\ts} E + F) + g (\p_{\ts}^2 E + \p_{\ts} F + G)
\ee
and respectively $f(\s)$, for which
\be
f'' E + f' (2\p_\s E -F) + f (\p_\s^2 E - \p_\s F + G)
\ee
Two distributions are equivalent if the terms multiplying $f$, $g$ and their  various  derivatives are the same . \label{equivfoot}}.
At the end of the day, we find the very simple constraint 

\be
\{ \H_R, \tilde A\} - \{ \tilde \H_R, A\} + \frac{A \d'}{\sqrt{}} + \frac{\tilde A \d'}{\sqrt{\widetilde{}}} =0
\ee
Let us now write  $A= A_0 + \hat A$, where $A_0 \equiv \p_\s \left(\frac{v}{R_v} \frac{\H_R}{\sqrt{}} \right) $. 
Evaluating
\be
R_v\left( \{ \H_R, A_0\} + \frac{ A_0 \d'}{\sqrt{}} \right) =  \frac{2 v \H_R^{(0)}}{\sqrt{}^3} \d'' + \frac{\H_R}{\sqrt{}} \d' + 2 \p_\s \left(v \frac{\H_R^{(0)}}{\sqrt{}^3} \right) \d'
\ee
the constraint we obtain on $\hat A$ is then, simply 

\be
\{ \H_R, \tilde{\hat A}\} - \{ \tilde \H_R, \hat A\} + \frac{\hat A \d'}{\sqrt{}} + \frac{\tilde{\hat A} \d'}{\sqrt{\widetilde{}}} = - \frac{1}{R_v} \left( \frac{\H_R}{\sqrt{}} + \frac{\tilde \H_R}{\sqrt{\tilde{}}}\right) \d'
\ee
This is solved by $\hat A = \H_R/R_v + \ldots$, where the $\ldots$ are periodic solutions to the homogenous equation, such as the  $a\p_\s \H_R/\sqrt{}$ term. 

Note however that for this value of $\hat A$, we need to revisit the conservation equation \eqref{conseqA}, which receives a new contribution from

\be
\int d\ts \{ A_p (\ts, \H_R(\s)\} = \frac{1}{R_v} \{ E_R, \H_R\} = \frac{1}{R_v} \p_\s \frac{\H_R}{\sqrt{}}
\ee
This contribution is very easy to cancel by including an explicit time-dependent term, $t/R_v \p_\s \H_R/\sqrt{}$.  This term is also consistent with the $K_U$ and $\H_R$ Jacobi identities. Therefore, the final solution we find for the integration function $A$ is

\be
\boxed{A = \p_\s\left[ \left( \frac{v}{R_v} +a\right)\frac{\H_R}{\sqrt{}}\right] + \frac{1}{R_v} \left( \H_R + t \p_\s \frac{\H_R}{\sqrt{}}\right)=\p_\s\left[ \left( \frac{v+t}{R_v} +a\right)\frac{\H_R}{\sqrt{}}\right]  + \frac{\H_R}{R_v}} \label{solnA}
\ee
which has the nice feature of not being explicitly time-dependent, since $v+t = \s - \l \phi$.

\subsection{Poisson bracket of $\chi$ with $\J_-$}

Another  commutator that requires special attention is that of $\chi$ with $\J_-$. Integrating the $\{\H_R, \tilde \J_-\}$ commutator we obtain
\be
\{\chi, \tilde \J_- \}= - \frac{\J_-}{\sqrt{}} \d(\s-\ts) + \p_{\ts}  \frac{\tilde \J_-}{\sqrt{\tilde{}}} \, \Theta(\s-\ts) +B(\ts)
\ee
In order for the commutator with $\bar J_0$ to be independent of the starting point of the interval, we need the winding of $B$ to equal $\J_-(0)/\sqrt{}(0)$. $B$ should also satisfy all the relevant Jacobi identities. 

As before, we first look at the  time derivative of this commutator, and try to fix $B$ by requiring that the Jacobi identity hold. We make use of 

\be
\{ H, \J_-/\sqrt{}\} = \frac{1+\l K_U}{1-\l K_U} \p_\s \frac{\J_-}{\sqrt{}}
\ee
and find that $B$ satisfies 
\be
\p_t B - \{ H, B\} = - \p_\s \left(\frac{1+\l K_U}{1-\l K_U} B \right) + \frac{1}{R_v} \p_\s \frac{\J_-}{\sqrt{}}
\ee
There are several solutions to this equation that have the correct winding, such as
\be
 B(\s) = \p_\s \left( \frac{\s}{R} \, \frac{\J_-}{\sqrt{}}\right) + b \, \p_\s \frac{\J_-}{\sqrt{}} + \frac{t}{R_v} \p_\s \frac{\J_-}{\sqrt{}}
\ee
 or 
\be
B(\s) = \p_\s \left( \frac{v}{R_v} \frac{\J_-}{\sqrt{}}\right) + b \, \p_\s \frac{\J_-}{\sqrt{}} + \frac{t}{R_v} \p_\s \frac{\J_-}{\sqrt{}}
\ee
for some constant $b$. To find which solution is correct, we need to analyse some further Jacobi identities.

\subsubsection*{Constraint from the commutator with $K_U$ }

 We first check the Jacobi identity for $\{ K_U, \{ \chi_0, \tilde \J_-\} \}$. The consistency condition we obtain is  
\be
 \{ K_U, (R-\ts) \p_{\ts} \frac{\tilde{\J}_-}{\sqrt{\tilde{}}}  + R \tilde B\} + \frac{\l \d'}{2 \sqrt{}} \left[(R-\s ) \p_\s \frac{\J_-}{\sqrt{}} + R B \right] = \frac{\l}{2\sqrt{}} \p_\s \frac{\J_-}{\sqrt{}} \d(\s-\ts) 
\ee
Using the fact that 
\be
\{ K_U , \frac{\tilde \J_-}{\sqrt{\tilde{}}}\} =\frac{\l}{2\sqrt{}} \p_\s \frac{\J_-}{\sqrt{}} \d(\s-\ts) 
\ee
and the criterion  for the equivalence of two distributions, we can reduce the above equation to 

\be
\{ K_U, \tilde B\} + \frac{\l \d'}{2 \sqrt{}}  B =0
\ee
It is easy to see that the terms proportional to $\p_\s  \J_-/\sqrt{}$ simply drop out of this equation. We then check that the first Ansatz does not solve this equation, whereas the second one does.
% As intermediate steps, we used  \emph{Correct?}
%%
%\be
%(R-\s ) \p_\s \frac{\J_-}{\sqrt{}} + R B = R \p_\s \frac{\J_-}{\sqrt{}} (1-\frac{t}{R_v}) - \frac{\l R}{R_v} \p_\s \left(\hat \phi \frac{\J_-}{\sqrt{}} \right)+ \frac{ \J_-}{ \sqrt{}}
%\ee
%
%\be
%\{ K_U, \tilde{\hat{\phi}}\frac{\tilde \J_-}{\sqrt{\tilde{}}} \} = \frac{\l \d(\s-\ts)}{2 \sqrt{}} \p_\s \left(\hat \phi \frac{\J_-}{\sqrt{}} \right) - \frac{R_v \J_- }{2 R (\sqrt{})^2}\d(\s-\ts)
%\ee
%The commutator with the zero mode reads
%\be
% \{ \chi_0, \J_-\} =R \p_{\s} \left( \frac{{\J}_-}{\sqrt{}} +\hat B (\s)\right)\;, \;\;\;\;\;\;\; \hat B = \left( \frac{v+t}{R_v} - \frac{\s}{R}\right) \frac{\J_-}{\sqrt{}} + b \frac{\J_-}{\sqrt{}}
%\ee
%Note that $\hat B$ does not have winding. It is also easy to show that  $ \{\chi, \bar J_0\} =0$.

\subsubsection*{Constraint from the $\H_R$ commutator}

The Jacobi identity reads

\be
\{ \H_R, \{ \chi_0, \tilde \J_-\}\} + \{ \{ \chi_0, \H_R\},\tilde \J_-\} - \{ \chi_0, \{ \H_R, \tilde \J_-\}\} =0
\ee
and from its form, it is easy to see it will relate the two integration functions $A$ and $B$. To simplify the constraint, we use

\be
\{ \H_R, \frac{ \tilde \J_-}{\sqrt{\tilde{}}} \} = - \frac{\J_- (1-\l \J_+)}{\sqrt{}^3} \d' + \left( \frac{1}{\sqrt{}} \p_\s \frac{\J_-}{\sqrt{}}  - \p_\s \frac{\J_- (1-\l \J_+)}{\sqrt{}^3}\right) \d(\s-\ts)
\ee

\be
\{ \frac{\H_R}{\sqrt{}}, \tilde \J_-\} = - \frac{\J_- (1-\l \J_+)}{\sqrt{}^3} \d'
\ee
and  find that $A$ and $B$ must satisfy the simple relation
\be
\{ \H_R, \tilde B\} + \{ A, \tilde \J_-\} + \frac{B}{\sqrt{}} \d'=0
\ee
 Letting $A=A_0+\hat A$, $B=B_0 +\hat B$ with $A_0 = \p_\s (v \H_R/ R_v \sqrt{})$ and $B_0 = \p_\s (v \J_-/R_v \sqrt{})$, we find the following constraint
\be
\{ \H_R, \tilde{\hat B}\} + \{ \hat A, \tilde \J_-\} + \frac{\hat B}{\sqrt{}} \d'= \frac{\l}{2R_v}\frac{ \tilde \H_R}{ \sqrt{\tilde{}}} \d' - \frac{1}{R_v} \frac{\J_-}{\sqrt{}} \d'
\ee
The already known $\H_R/R_v$ contribution to $\hat A$ accounts for the last term on the right-hand side. However, we need a new term in $\hat B$ to account for the first term.  It is clear that 
\be
\hat B = - \frac{\l \H_R}{2R_v}
\ee
 does the job, and one can check that it is also consistent with the previous consistency conditions. 
 
Finally,  it is not hard to check that any term in $B$ proportional to $\p_\s \J_-/\sqrt{}$ and in $A$ with  $\p_\s \H_R/\sqrt{}$ (with the same proportionality coefficient) automatically satisfies this equation. This sets $a=b$. We can also check the time dependence matches exactly. 

To summarize, the solution that we have found for B that is consistent with all the Jacobi identities we have checked is
\be
\boxed{
B(\s) = \p_\s \left[\left( \frac{v+t}{R_v} +a\right)\frac{\J_-}{\sqrt{}}\right] - \frac{\l \H_R}{2R_v}}
\ee
for the same arbitrary constant $a$ as in \eqref{solnA}.

\subsection{Other commutators}

The Poisson brackets of $\chi$ with all the remaining fields, such as $\J_+$ or $\P$, are determined by its commutators with $\H_R$ and $\J_-$ (i.e., the functions $A,B$) and its commutators  with $\H_L = \H_R+\P$ and $K_U=\J_++\l \H_R/2$, which are local. We thus find

%
% It follows that
%%
%\be
%\left\{  \chi_0,  \H_R \right\} = -\frac{ \H_R}{\sqrt{}} +  R \, \p_{\s} \left[\frac{\H_R}{\sqrt{}} (a+1- \frac{\l \hat{\phi}}{R_v})\right]  + \frac{R}{R_v} \H_R\equiv-\frac{ \H_R}{\sqrt{}} + \frac{R}{R_v} \H_R +R\, \p_\s  \left(\frac{\H_R}{\sqrt{}}  +\hat A \right)
%\ee
%which implies the expected commutator of $\chi_0$ with $H_R$. Next, the commutator with $\H_L = \H_R + \P$ is 
%
%\be
%\{\H_L, \tilde \chi\} = \left(\frac{\H_R}{\sqrt{}} -\H_R \right) \d(\s-\ts)
%\ee
%to which we could, in principle, add an arbitrary functin of $\s$, but it would be very unnatural to do so, so we drop it. From this, we deduce that the commutator with the momentum reads

\be
\{ \P, \tilde \chi\} = - \left(\H_R + \frac{\H_R}{\sqrt{}}\right) \d(\s-\ts) + \p_\s \frac{\H_R}{\sqrt{}} \, \Theta(\ts-\s) + A(\s)
\ee
%We thus find
%\be
%\left\{ \chi_0, \P\right\}=\H_R -\frac{R}{R_v} \H_R -R\, \p_\s  \left(\frac{\H_R}{\sqrt{}}  +\hat A \right) = - \frac{\l w}{R_v} \H_R -R\, \p_\s  \left(\frac{\H_R}{\sqrt{}}  +\hat A \right)
%\ee
%with $\hat A= (a - \frac{\l \hat \phi}{R_v} ) \H_R/\sqrt{}$. From the $
%\{\chi, \tilde K_U\} $ commutator above, 
%it  follows that 
%%
\be
\{\chi, \tilde \J_+\} = \frac{\l \H_R}{2\sqrt{}} \d(\s-\ts) - \frac{\l}{2} \p_{\ts} \frac{\tilde{\H}_R}{\sqrt{\tilde{} }} \, \Theta(\s-\ts) - \frac{\l}{2} A(\ts)
\ee
%and so
%\be
% \{ \chi_0, \J_+\} = - \frac{\l R}{2} \p_\s\left( \frac{\H_R}{\sqrt{}} + \hat A\right) - \frac{\l R}{2 R_v} \H_R
%\ee
Combining the latter with the $\{\chi, \tilde \J_-\}$ commutator and integrating, we find the $\{\chi, \tilde \phi\}$ commutator

\be
\{ \chi, \tilde \phi\} = - \frac{\tilde{\J}_-+ \l \tilde \H_R/2}{\sqrt{\tilde{}}} \left( \Theta (\s-\ts) + \frac{\tilde v+\tilde t}{R_v} +a \right)  + C(\s)
\ee
The $\{ \H_R, \tilde \phi\} $ commutator requires that $C'(\s) =0$, so $C = c $, a constant, %The Jacobi identity for the $\{ K_U,\{ \chi_0,\tilde \phi\}\}$ commutator yields the constraint $\{ K_U, c \} =0$, and does not fix $a$, as long as it is a constant. % . Possibly $c=0$ is a safe choice, but we will try to avoid assuming this.
%\emph{Check also commutator with $\H_R$!}
which we will set to zero.

\section{Flow equations for the currents and the charges \label{flowapp}}

Let $\Dl = \p_\l + \{ \O_{tot}, \cdot\}$, where for the purposes of computing commutators, the expression

\be
\O_{tot} = \frac{w \chi_0}{R} - \int d\s \H_R \hat \phi
\ee 
is significantly easier to use. Using the commutators with the zero modes, we find that at \emph{classical } level, the currents satisfy the following flow equations

\be
\Dl K_U = \Dl \H_L =0 \label{flowleftcurr}
\ee
The equation for $\H_R$ is  
\be
\Dl \H_R = \frac{w }{R_v} \H_R +  \p_\s \left[\frac{\H_R}{\sqrt{}} \left( w(1+a)-\frac{R \hat \phi}{R_v}\right)\right]  \label{flowHR}
\ee
which implies the following flow equation for $\chi$

\be
\mathcal{D}_\l \chi = \frac{\H_R}{\sqrt{}}  \left( w(1+a)-\frac{R \hat \phi}{R_v}\right)  + \frac{w}{R_v} \chi
\ee
The flow equation for $\J_-$ is given by

\be
\mathcal{D}_\l \J_- = - \frac{1}{2 } \H_R  - \frac{w\l }{2R_v} \H_R + \p_\s \left[\frac{\J_-}{\sqrt{}} \left( w(1+a)-\frac{R \hat \phi}{R_v}\right) \right]%= - \frac{1}{2 } \H_R  - \p_\s \left[\frac{\J_-}{\sqrt{}} \frac{ R \hat \phi}{R_v} \right]- \frac{w \l R}{2R_v} \H_R 
\ee
which, together with \eqref{flowHR}, implies that 
\be
\mathcal{D}_\l  (\J_- + \frac{\l}{2} \H_R)  =  \p_\s \left[\frac{\J_-+ \l \H_R/2}{\sqrt{}} \left( w(1+a)-\frac{R \hat \phi}{R_v}\right) \right]
\ee
Finally, the flow equation for $\phi$ is
\be
\mathcal{D}_\l \phi(\s) = %\frac{\J_- + \l \H_R/2 }{\sqrt{}} \hat \phi(\s ) + \frac{w}{R} \{ \chi_0, \phi(\s)\} = 
- \frac{\J_- + \l \H_R/2 }{\sqrt{}} \left( w(1+a)-\frac{R \hat \phi}{R_v}\right)
\ee
We subsequently use these flow equations to compute the flow of the conserved charges. We trivially have $\Dl Q_f = \Dl P_\eta =0$. As for the right-moving charges, we obtain
\be
\Dl \bar Q_{\bar f} = \frac{w}{R_v} \bar Q_{\bar f} - \frac{w}{R_v} \left(a+1 + \frac{t}{R_v}\right) \bar Q_{\bar f'}\;, \;\;\;\;\;\;\; \Dl \bar P^{KM}_{\bar \eta} =  - \frac{w}{R_v} \left(a+1 + \frac{t}{R_v}\right) \bar P^{KM}_{\bar \eta'}
\ee


\begin{thebibliography}{99}
\bibitem{Maldacena:1997re}
J.~M.~Maldacena,
``The Large N limit of superconformal field theories and supergravity,''
Int. J. Theor. Phys. \textbf{38} (1999), 1113-1133
%doi:10.1023/A:1026654312961
arXiv:hep-th/9711200 [hep-th].

\bibitem{Li:2010dr}
W.~Li and T.~Takayanagi,
``Holography and Entanglement in Flat Spacetime,''
Phys. Rev. Lett. \textbf{106} (2011), 141301
%doi:10.1103/PhysRevLett.106.141301
arXiv:1010.3700 [hep-th].

\bibitem{ElShowk:2011cm}
S.~El-Showk and M.~Guica,
``Kerr/CFT, dipole theories and nonrelativistic CFTs,''
JHEP \textbf{12} (2012), 009
%doi:10.1007/JHEP12(2012)009
arXiv:1108.6091 [hep-th].

\bibitem{Smirnov:2016lqw}
F.~A.~Smirnov and A.~B.~Zamolodchikov,
``On space of integrable quantum field theories,''
Nucl. Phys. B \textbf{915} (2017), 363-383
%doi:10.1016/j.nuclphysb.2016.12.014
arXiv:1608.05499 [hep-th].

\bibitem{Cavaglia:2016oda}
A.~Cavagli\`a, S.~Negro, I.~M.~Sz\'ecs\'enyi and R.~Tateo,
``$T \bar{T}$-deformed 2D Quantum Field Theories,''
JHEP \textbf{10} (2016), 112
%doi:10.1007/JHEP10(2016)112
arXiv:1608.05534 [hep-th].

\bibitem{Dubovsky:2012wk}
S.~Dubovsky, R.~Flauger and V.~Gorbenko,
``Solving the Simplest Theory of Quantum Gravity,''
JHEP \textbf{09} (2012), 133
%doi:10.1007/JHEP09(2012)133
arXiv:1205.6805 [hep-th].

\bibitem{Dubovsky:2013ira}
S.~Dubovsky, V.~Gorbenko and M.~Mirbabayi,
``Natural Tuning: Towards A Proof of Concept,''
JHEP \textbf{09} (2013), 045
%doi:10.1007/JHEP09(2013)045
arXiv:1305.6939 [hep-th].

\bibitem{Cardy:2018sdv}
J.~Cardy,
``The $ T\overline{T} $ deformation of quantum field theory as random geometry,''
JHEP \textbf{10} (2018), 186
%doi:10.1007/JHEP10(2018)186
arXiv:1801.06895 [hep-th].

\bibitem{Dubovsky:2018bmo}
S.~Dubovsky, V.~Gorbenko and G.~Hern\'andez-Chifflet,
``$ T\overline{T} $ partition function from topological gravity,''
JHEP \textbf{09} (2018), 158
%doi:10.1007/JHEP09(2018)158
arXiv:1805.07386 [hep-th].

\bibitem{Cardy:2019qao}
J.~Cardy,
``$T\bar T$ deformation of correlation functions,''
JHEP \textbf{19} (2020), 160
%doi:10.1007/JHEP12(2019)160
arXiv:1907.03394 [hep-th].

\bibitem{Kruthoff:2020hsi}
J.~Kruthoff and O.~Parrikar,
``On the flow of states under $T\overline{T}$,''
arXiv:2006.03054 [hep-th].

\bibitem{Giveon:2017nie}
A.~Giveon, N.~Itzhaki and D.~Kutasov,
``$ \mathrm{T}\overline{\mathrm{T}} $ and LST,''
JHEP \textbf{07} (2017), 122
%doi:10.1007/JHEP07(2017)122
arXiv:1701.05576[hep-th].


\bibitem{Apolo:2018qpq}
L.~Apolo and W.~Song,
``Strings on warped AdS$_{3}$ via $ \mathrm{T}\bar{\mathrm{J}} $ deformations,''
JHEP \textbf{10} (2018), 165
%doi:10.1007/JHEP10(2018)165
arXiv:1806.10127 [hep-th].

\bibitem{Chakraborty:2018vja}
S.~Chakraborty, A.~Giveon and D.~Kutasov,
``$ J\overline{T} $ deformed CFT$_{2}$ and string theory,''
JHEP \textbf{10} (2018), 057
%doi:10.1007/JHEP10(2018)057
arXiv:1806.09667 [hep-th].


\bibitem{Guica:2017lia}
M.~Guica,
``An integrable Lorentz-breaking deformation of two-dimensional CFTs,''
SciPost Phys. \textbf{5} (2018) no.5, 048
%doi:10.21468/SciPostPhys.5.5.048
arXiv:1710.08415 [hep-th].

\bibitem{Anous:2019osb}
T.~Anous and M.~Guica,
``A general definition of $JT_a$ -- deformed QFTs,''
arXiv:1911.02031 [hep-th].

\bibitem{Frolov:2019xzi}
S.~Frolov,
``$T{\overline T}$, $\widetilde JJ$, $JT$ and $\widetilde JT$ deformations,''
J. Phys. A \textbf{53} (2020) no.2, 025401
%doi:10.1088/1751-8121/ab581b
arXiv:1907.12117 [hep-th].

\bibitem{LeFloch:2019rut}
B.~Le Floch and M.~Mezei,
``Solving a family of $T\bar{T}$-like theories,''
arXiv:1903.07606 [hep-th].

\bibitem{Guica:2020uhm}
M.~Guica and R.~Monten,
``Infinite pseudo-conformal symmetries of classical $T \bar T$, $J \bar T $ and $J T_a$ - deformed CFTs,''
arXiv:2011.05445 [hep-th].

\bibitem{Guica:2019nzm}
M.~Guica and R.~Monten,
``$T\bar T$ and the mirage of a bulk cutoff,''
arXiv:1906.11251 [hep-th].

\bibitem{Bzowski:2018pcy}
A.~Bzowski and M.~Guica,
``The holographic interpretation of $J \bar T$-deformed CFTs,''
JHEP \textbf{01} (2019), 198
%doi:10.1007/JHEP01(2019)198
arXiv:1803.09753 [hep-th].

\bibitem{Schwimmer:1986mf}
A.~Schwimmer and N.~Seiberg,
``Comments on the N=2, N=3, N=4 Superconformal Algebras in Two-Dimensions,''
Phys. Lett. B \textbf{184} (1987), 191-196
%doi:10.1016/0370-2693(87)90566-1

\bibitem{Guica:2019vnb}
M.~Guica,
``On correlation functions in $J\bar T$-deformed CFTs,''
J. Phys. A \textbf{52} (2019) no.18, 184003
%doi:10.1088/1751-8121/ab0ef3
arXiv:1902.01434 [hep-th].

%\bibitem{mirage} Mirage paper
%\bibitem{cardy} Cardy correlators
%\bibitem{kutbeyond} Kutasov beyond Ads
%\bibitem{rosenhaus} Rosenhaus and Smolkin
%\bibitem{liumar} Liu, Kraus, Marolf
%\bibitem{jtcorr} $J\bar T$ correlators
%\bibitem{zamtba} Zamolodchikov TBA
%\bibitem{jtbar} first $J\bar T$ paper
%\bibitem{cssint} holography for $J\bar T$
\end{thebibliography}
\end{document}